\documentclass[prr,aps,reprint,a4paper,superscriptaddress,floatfix,longbibliography]{revtex4-2}
\usepackage{times}
\usepackage{graphicx}
\usepackage{epsfig}
\usepackage{amsfonts}
\usepackage{amsmath}
\usepackage{amssymb}
\usepackage{dsfont}
\usepackage{amsthm}
\usepackage{color}
\usepackage[colorlinks=true,linkcolor=blue,citecolor=blue,urlcolor=blue]{hyperref}
\usepackage{comment}
\usepackage{braket}
\usepackage{physics}
\usepackage{multirow}
\usepackage{float}

\newtheorem{theorem}{Theorem}

\newtheorem{definition}{Definition}

\newtheorem{observation}{Observation}

\usepackage[ruled,linesnumbered]{algorithm2e}
\usepackage{mathdots}
\usepackage{MnSymbol}
\usepackage{diagbox}
\usepackage{enumitem} 

\SetCommentSty{mycommfont}

\begin{document}


\title{Equivalence between face nonsignaling correlations, full nonlocality, all-versus-nothing proofs, and pseudotelepathy}


\author{Yuan Liu}
\affiliation{School of Computing and Data Science, The University of Hong Kong, Pokfulam Road, 999077 Hong Kong, China}

\author{Ho Yiu Chung}
\affiliation{School of Computing and Data Science, The University of Hong Kong, Pokfulam Road, 999077 Hong Kong, China}

\author{Emmanuel Zambrini Cruzeiro}
\affiliation{Instituto de Telecomunicaç\~{o}es, 1049-001 Lisbon, Portugal}

\author{Junior R. Gonzales-Ureta}
\affiliation{Department of Physics, Stockholm University, 10691 Stockholm, Sweden}

\author{Ravishankar Ramanathan}
\email{ravi@cs.hku.hk}
\affiliation{School of Computing and Data Science, The University of Hong Kong, Pokfulam Road, 999077 Hong Kong, China}

\author{Ad\'an~Cabello}
\email{adan@us.es}
\affiliation{Departamento de F\'{\i}sica Aplicada II, Universidad de Sevilla, E-41012 Sevilla,
Spain}
\affiliation{Instituto Carlos~I de F\'{\i}sica Te\'orica y Computacional, Universidad de
Sevilla, E-41012 Sevilla, Spain}


\begin{abstract}
We show that a quantum correlation $p$ is in a face of the nonsignaling polytope with no local points if and only if $p$ has nonlocal content~$1$, if and only if $p$ allows for a Greenberger-Horne-Zeilinger-like proof, and if and only if $p$ provides a perfect strategy for a nonlocal game. That is, face nonsignaling (FNS) correlations, full nonlocality (FN), all-versus nothing (AVN) proofs, and pseudotelepathy (PT) are equivalent. This shows that different resources behind a wide variety of fundamental results are in fact the same resource. We demonstrate that quantum correlations with FNS=FN=AVN=PT do not need to maximally violate a tight Bell inequality. We introduce a method for identifying quantum FNS=FN=AVN=PT correlations and use it to prove quantum mechanics does not allow for FNS=FN=AVN=PT neither in the $(3,3;3,2)$ nor in the $(3,2;3,4)$ Bell scenarios. This solves an open problem that, because of the FNS=FN=AVN=PT equivalence, has implications in several fields.
\end{abstract}


\maketitle


{\em Introduction.} Bell nonlocality, i.e., the violation of Bell inequalities \cite{Bell:1964PHY}, is one of the most fundamental predictions of quantum mechanics (QM). Quantum nonlocal correlations have a wide scope of applications, ranging from secure communication \cite{Ekert:1991PRL} and randomness amplification \cite{Colbeck:2006XXX} to reduction of communication complexity \cite{Brukner:2004PRL} and self-testing of quantum devices \cite{Yao_self}.
However, not all Bell nonlocal correlations are equally powerful. Specific tasks require specific types of correlations. Hereafter, we will focus on four types that have attracted interest for different reasons.

\begin{figure}[t!]
\includegraphics[width=8.7cm]{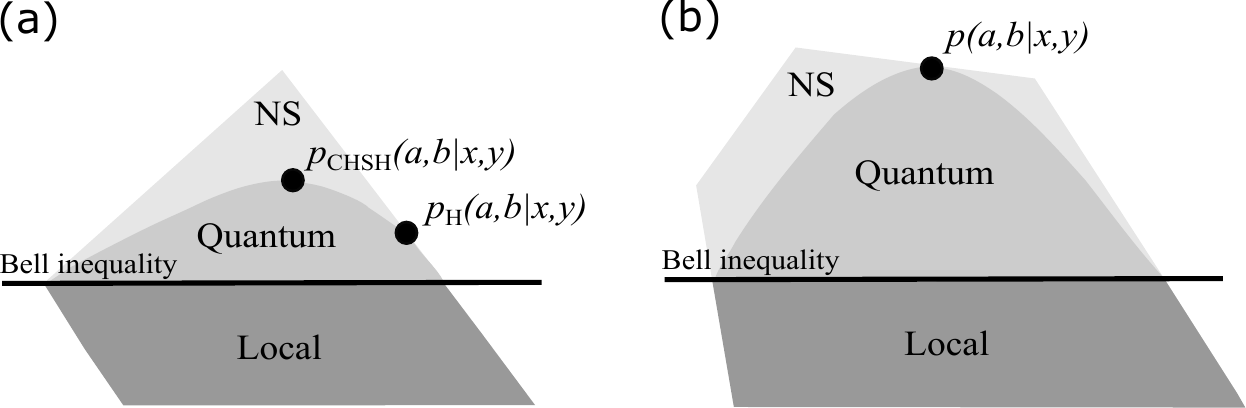}
\caption{(a) In the $(2,2;2,2)$ Bell scenario, $p_{\rm CHSH}(a,b|x,y)$ is the quantum correlations that maximally violates the Clauser-Horne-Shimony-Holt (CHSH) Bell inequality \cite{Clauser:1969PRL}. $p_{\rm CHSH}(a,b|x,y)$ is far from the faces of the NS polytope. $p_{\rm H}(a,b|x,y)$ is the correlation that corresponds to the optimal proof of Bell nonlocality of Hardy \cite{HardyPRL1993}. $p_{\rm H}(a,b|x,y)$ in a face of the NS polytope that contains a local point \cite{Goh2018PRA}. (b)~In the $(3,4;3,4)$ Bell scenario, there is a correlation $p(a,b|x,y)$ \cite{Cabello:2001PRLa,Cabello:2001PRLb} that is in a face of the NS polytope that has no local points \cite{Aolita:2012PRA}.} 
\label{fig_FN_01}
\end{figure}

{\em Face nonsignaling correlations.} Consider a bipartite Bell scenario $(|X|,|A|;|Y|,|B|)$, where $x \in X$ and $y \in Y$ are Alice's and Bob's measurement settings, respectively, and $a \in A$ and $b \in B$ are Alice's and Bob's measurement outcomes. A Bell nonlocal correlation $p(a,b|x,y)$ in this scenario is a point outside the set of local correlations (the local polytope) and inside the set of correlations satisfying nonsignaling (NS) (the NS polytope) \cite{Pitowsky:1989}. Using the results in \cite{Cabello:2010XXX,CSWPRL2014}, it can be proven that neither QM \cite{Ramanathan:2010PRL} nor any theory that assigns probabilities to sharp observables
can attain a nonlocal {\em vertex} of a NS polytope. Still, quantum Bell nonlocal correlations can be in a face of the NS polytope. There are two possibilities \cite{Goh2018PRA}: either the Bell nonlocal correlation is in a face that contains local points \cite{Goh2018PRA}, see Fig.~\ref{fig_FN_01}(a), or it is {\em in a face that does not contain local points}, see Fig.~\ref{fig_FN_01} (b). The Bell nonlocal correlations of the second type are called face nonsignaling (FNS). Why are FNS correlations so important? Some reasons are the following:

(i) Most quantum Bell nonlocal correlations can be classically simulated either by relaxing the assumption of measurement independence \cite{Shimony:1993} (and admitting that some measurement settings may depend on hidden variables) or by relaxing the assumption of parameter independence \cite{Shimony:1993} (and admitting that some outcomes may depend on spacelike separated settings). However, there are quantum Bell nonlocal correlations that {\em cannot} be classically simulated unless both assumptions are totally removed \cite{Vieira:2024X}. It can be proven that these Bell nonlocal correlations {\em must} be FNS or arbitrarily close to it \cite{Vieira:2024X}. 

(ii) FNS correlations are fundamental for identifying the principle that bounds quantum correlations \cite{Popescu:1994FPH}, since
a way to obtain the quantum bounds is by noticing that the studied scenario may be naturally linked to a larger scenario in which QM allows for FNS correlations that imply bounds on the smaller scenario \cite{Cabello:2015PRL,Cabello:2019PRA}.

(iii) FNS correlations opened up the possibility of perfect randomness from seeds with arbitrarily weak randomness \cite{Gallego:2013NC}. 


{\em Full nonlocality.} One of the most widely used measures of Bell nonlocality is the nonlocal content \cite{Elitzur:1992PLA}. Given a NS correlation $p(a,b|x,y)$, consider all possible decompositions of the form
\begin{equation}
\label{deco}
p(a,b|x,y) = q_{\mathrm{L}} p_{\mathrm{L}}(a,b|x,y) + (1 - q_{\mathrm{L}})p_{\mathrm{NL}}(a,b|x,y),
\end{equation}
in terms of local correlations $p_{\mathrm{L}}(a,b|x,y)$ and Bell nonlocal NS correlations $p_{\mathrm{NL}}(a,b|x,y)$, with respective weights $q_{\mathrm{L}}$ and $1 - q_{\mathrm{L}}$, with $0\leq q_{\mathrm{L}} \leq 1$. The local content or local fraction $q_{\mathrm{L}}^{\mathrm{max}}$ of $p(a,b|x,y)$ is the maximum local weight over all decompositions of the form (\ref{deco}). That is, $q_{\mathrm{L}}^{\mathrm{max}} \doteq \max_{\{p_{\mathrm{L}},p_{\mathrm{NL}}\}} q_{\mathrm{L}}$. The nonlocal content is $q_{\mathrm{NL}}^{\mathrm{min}} \doteq 1-q_{\mathrm{L}}^{\mathrm{max}}$.
The correlation is local if and only if $q_{\mathrm{NL}}^{\mathrm{min}}=0$. 
The correlation is said to have {\em full nonlocality} (FN) \cite{Aolita:2012PRA} or {\em strong nonlocality} \cite{Abramsky2019Pro} if $q_{\mathrm{NL}}^{\mathrm{min}}=1$. For example, the maximum quantum violation of the CHSH Bell inequality \cite{Clauser:1969PRL} has $q_{\mathrm{NL}}^{\mathrm{min}}=\sqrt{2}-1\approx 0.414$.

FN is {\em necessary} for some quantum information tasks. For example: (I) Improving the number of classical messages that can be sent without error through a single use of a classical channel \cite{Cubitt:2010PRL}, (II) device-independent quantum key distribution (DI-QKD) based on perfect correlations \cite{Horodecki:2010XXX}, (III) DI-QKD based on the magic square game \cite{Zhen:2023PRL}, and (IV) some types of DI-QKD based on parallel repetition \cite{Vidick:2017XXX,Jain:2020IEE}.


{\em All-versus-nothing proofs.} The Einstein-Podolsky-Rosen \cite{Einstein:1935PR} argument suggesting the possibility of completing QM with local hidden variables (LHVs) was based on {\em perfect} correlations that allows the parties to predict {\em with certainty} the outcome of the measurement of a distant party. For this reason, the proofs of Bell's theorem of impossibility of LHVs that only use perfect correlations have a special status in foundations of QM. Examples of such proofs are: (i') the proofs of Stairs \cite{Stairs:1983PS} and Heywood and Redhead \cite{HR83} based on the Kochen-Specker theorem \cite{Kochen:1967JMM}, (ii') the proof of Greenberger, Horne, and Zeilinger (GHZ) with four parties \cite{GHZ89}, (iii') Mermin's simplification to three parties \cite{Mermin:1990AJP,Pan:2000NAT,Leibfried:2004Sci,Roos:2004Sci}, (iv') the ``all-versus-nothing'' (AVN) proof with two parties \cite{Cabello:2001PRLa,Cabello:2001PRLb,CinelliPRL2005,YangPRL2005,Aolita:2012PRA,Xu:2022PRL}, and (v') the proofs of Bell nonlocality based on stabilizers of graph states \cite{Guhne:2005PRL}. The name AVN was coined by Mermin \cite{Mermin:1990PRLa} to designate those proofs in which the conflict between QM and LHVs is evident by looking only at predictions with certainty.

In the bipartite case, AVN proofs can be characterized (see Appendix~A) as follows. A table of zeros for the $(|X|, |A|; |Y|, |B|)$ Bell scenario is a matrix with $|X| \times |A|$ rows and $|Y| \times |B|$ columns containing either zeros or empty entries. A zero in the entry $(a,b|x,y)$ indicates that the probability of $(a,b|x,y)$ is zero. An AVN proof consists of a quantum correlation that produces a table of zeros which cannot be realized by any LHV model. Given $S=S_A \cup S_B$, with $S_A = \{(a |x)\}_{x \in X, a \in A}$ and $S_B = \{(b |y)\}_{y \in Y, b \in B}$, a table of zeros is not realizable by any LHV if, for every assignment $f: S \rightarrow \{0,1\}$ satisfying $\sum_{a} f(a|x)=1$, $\forall x \in X$, and $\sum_{b} f(b|y)=1$, $\forall y \in Y$, there is a pair $\{(a|x),(b|y)\}$ for which $f(a|x)=f(b|y)=1$ and $p(a,b|x,y)=0$.


{\em Pseudotelepathy.} There is a form of Bell nonlocality that plays a fundamental role in quantum computation and quantum information. It is related to a specific type of nonlocal games. A bipartite nonlocal game \cite{CHTW04,Aravind:2004AJP,GBT05,BroadbentPhD2008} is a 4-tuple $G = (X \times Y, A \times B, \pi, W)$, where $X$ ($Y$) is the input set of the first player, Alice (the second player, Bob), $A$ ($B$) is the corresponding set of outputs, $\pi(X \times Y)$ is the distribution of inputs, and $W(X \times Y \times A \times B) \in \{0,1\}$ is the winning condition, i.e., the condition that inputs and outputs should satisfy to win the game. Consequently, the winning probability of the game is given by 
\begin{equation}
\label{win}
\omega(G) = \sum_{x,y,a,b} \pi(x,y) p(a,b|x,y) W(a,b,x,y).
\end{equation} 
The game $G$ admits a {\em perfect strategy} or {\em pseudotelepathy} (PT) \cite{GBT05,BroadbentPhD2008} if there is a correlation $p(a,b|x,y)$ that allows Alice and Bob to win every round of $G$. That is, if $W(a,b,x,y)=1$ for all $(a,b|x,y)$ such that $p(a,b|x,y)\neq 0$. Therefore, a quantum strategy offers PT whenever the quantum winning probability is $\omega_\mathrm{Q}(G)=1$, while using any classical strategy (that does not involve communication between Alice and Bob) the winning probability is $\omega_{\mathrm{C}}(G) < 1$.

PT is crucial in: (I') the proof of the quantum computational advantage for shallow circuits \cite{Bravyi:2018SCI}, (II') the proof of MIP$^\ast$=RE \cite{Ji:2021CACM}, (III') device-independent randomness generation in a network with untrusted users, since PT allows to certify more local randomness (randomness known to one party but not to the other) and in a more robust way than standard Bell nonlocality \cite{Fu:2018PRA}, (IV') multiple access channels: if two senders that aim to transmit individual messages to a single receiver have PT, then they can transmit information at the maximal possible rate \cite{LALS2020NATCOM}.


{\em Equivalence.} Our first result is the following.


\begin{theorem} \label{th1}
The following statements are equivalent:

(i) A quantum correlation $p$ is face nonsignaling.

(ii) $p$ has full nonlocality.

(iii) $p$ allows for an all-versus-nothing proof.

(iv) $p$ allows for a perfect (or pseudotelepathy) strategy.
\end{theorem}


{\em Proof.} The equivalence between (i) and (ii) follows from the fact that the nonlocal content $q_{\mathrm{NL}}^{\mathrm{min}}$ measures Bell nonlocality relative to the local and NS polytopes so that $q_{\mathrm{NL}}^{\mathrm{min}}$ takes the value $1$ if and only if $p$ is in a face that has no local points [since, otherwise, $q_{\mathrm{L}}^{\mathrm{max}}(p) \neq 0$; for example, for $p_\mathrm{H}$ in Fig.~\ref{fig_FN_01}~(a), $q_{\mathrm{NL}}^{\mathrm{min}} = 5\sqrt{5}-11 \approx 0.18$]
(see Appendix~B).

The equivalence between (iii) and (iv) follows from the observation that a quantum correlation $p$ yields quantum winning probability $1$ for the game $G$ in which the winning condition is achieving (in each context asked by the referee) all the zeros in the table of zeros of $p$, while the classical winning probability is strictly smaller than $1$, if and only if the table of zeros of $p$ cannot be realized by any LHV variable model (see Appendix~C).

The equivalence between (ii) [and (i)] and (iv) [and (iii)] can be proven as follows. 
To prove that (iv) implies (ii), let us observe that, by (iv), there is a game $G$ for which there is a quantum strategy (correlation) $p$ that provides a winning probability $\omega^{(p)}(G) = 1 = \omega_{\mathrm{NS}}(G)$, while $\omega_{\mathrm{C}}(G) < 1$, where $\omega_{\mathrm{NS}}(G)$ is the winning probability allowed by NS correlations.
Let us now consider any convex decomposition of $p$ of the form~\eqref{deco}. Then, by the linearity of the winning probability in Eq.~\eqref{win}, 
\begin{equation}
\omega^{(p)}(G) = q_{\mathrm{L}} \omega^{(p_{\mathrm{L}})}(G) + (1-q_{\mathrm{L}}) \omega^{(p_{\mathrm{NL}})}(G),
\end{equation}
where $\omega^{(p_{\mathrm{L}})}(G)$ and $\omega^{(p_{\mathrm{NL}})}(G)$ are the winning probabilities using the local correlation $p_{\mathrm{L}}$ and the NS correlation $p_{\mathrm{NL}}$, respectively. Since $\omega^{(p)}(G) = 1$ and $0 \leq \omega(G) \leq 1$, then $\omega^{(p_{\mathrm{L}})}(G) = 1$ whenever $0 < q_{\mathrm{L}}$. This contradicts the assumption that $\omega_{\mathrm{C}}(G) < 1$. Therefore, $q_{\mathrm{L}} = 0$ in any convex decomposition of $p$ of the form~\eqref{deco}. That is, $q_{\mathrm{L}}^{\mathrm{max}} = \max_{\{p_{\mathrm{L}},p_{\mathrm{NL}}\}} q_{\mathrm{L}} = 0$. 

To prove that (ii) implies (iv), let us observe that, by (ii), $q_{\mathrm{L}}^{\mathrm{max}}(p)=0$. As shown in \cite{Colbeck:2019PRA}, the local content can be computed by the following linear program: 
\begin{equation}
\label{fou}
\begin{aligned}
q_{\mathrm{L}}^{\mathrm{max}}(p) = \max_{} \quad & \sum_i q_i\\
\textrm{s.t.} \quad & \sum_i q_i P^{\mathrm{L}}_{i}\leq p\\
& q_i \geq 0 ~ \forall i. \\
\end{aligned}
\end{equation}
Here, $P^{\mathrm{L}}_i$ correspond to vertices of the local polytope and $\sum_i q_i P^{\mathrm{L}}_{i}\leq p$ must be interpreted term by term. The dual of this linear program can be written as follows:
\begin{equation}
\begin{aligned}
\min_{I} \quad & \tr(I^T p)\\
\textrm{s.t.} \quad & \tr(I^T P^{\mathrm{L}}_i) \geq 1 ~ \forall i~\\
& I \geq 0. \\
\end{aligned}
\end{equation}
Again, $I \geq 0$ must be interpreted term by term, and $i$ runs over all vertices of the local polytope.
By the strong duality theorem of linear programming \cite{Matouek:2006}, the dual and primal optima are equal when one of the two problems has an optimal solution [we have that $\min_I \tr(I^T p) = q_{\mathrm{L}}^{\mathrm{max}}(p) = 0$]. In other words, $I$ defines a Bell expression $\tr(I^T p)$ whose minimum value in QM is the algebraic minimum $0$ achieved by $p$, and whose minimum local value is $\geq 1$. 
Moreover, in order to achieve the algebraic minimum, the Bell expression $I(a,b,x,y)$ has to have coefficients equal to zero for every $p(a,b|x,y)>0$. This allows us to reformulate the Bell inequality for $I(a,b,x,y)$ as a nonlocal game $G$ with a PT strategy. The winning condition of $G$ is
\begin{equation}
W(a,b,x,y)= \left\{
\begin{array}{lr} 
1, & \text{if\;} I(a,b,x,y) = 0 \\
0, & \text{otherwise.} 
\end{array}
\right.
\end{equation}
That is, by taking the complement of $I$, we obtain the game $G$ with $\omega_{\mathrm{C}}(G) < 1$ and for which $p$ provides $\omega^{(p)}(G) = 1$. \hfill \qedsymbol


{\em FNS=FN=AVN=PT and Bell inequalities.} Gisin, M\'ethot, and Scarani~\cite{Gisin:2007IJQI} made the observation that all known quantum PT strategies correspond to maximum violations of tight Bell inequalities. They also raised the question of whether this is always the case. Our Theorem \ref{th1} shows that this is in fact an important question, as it does not only concern PT. Our second result answers the question in its more general version.


\begin{observation} \label{th2}
Not all quantum FNS=FN=AVN=PT correlations define tight Bell inequalities.
\end{observation}

\begin{figure}[t!]
\centering
\includegraphics[scale=0.14]{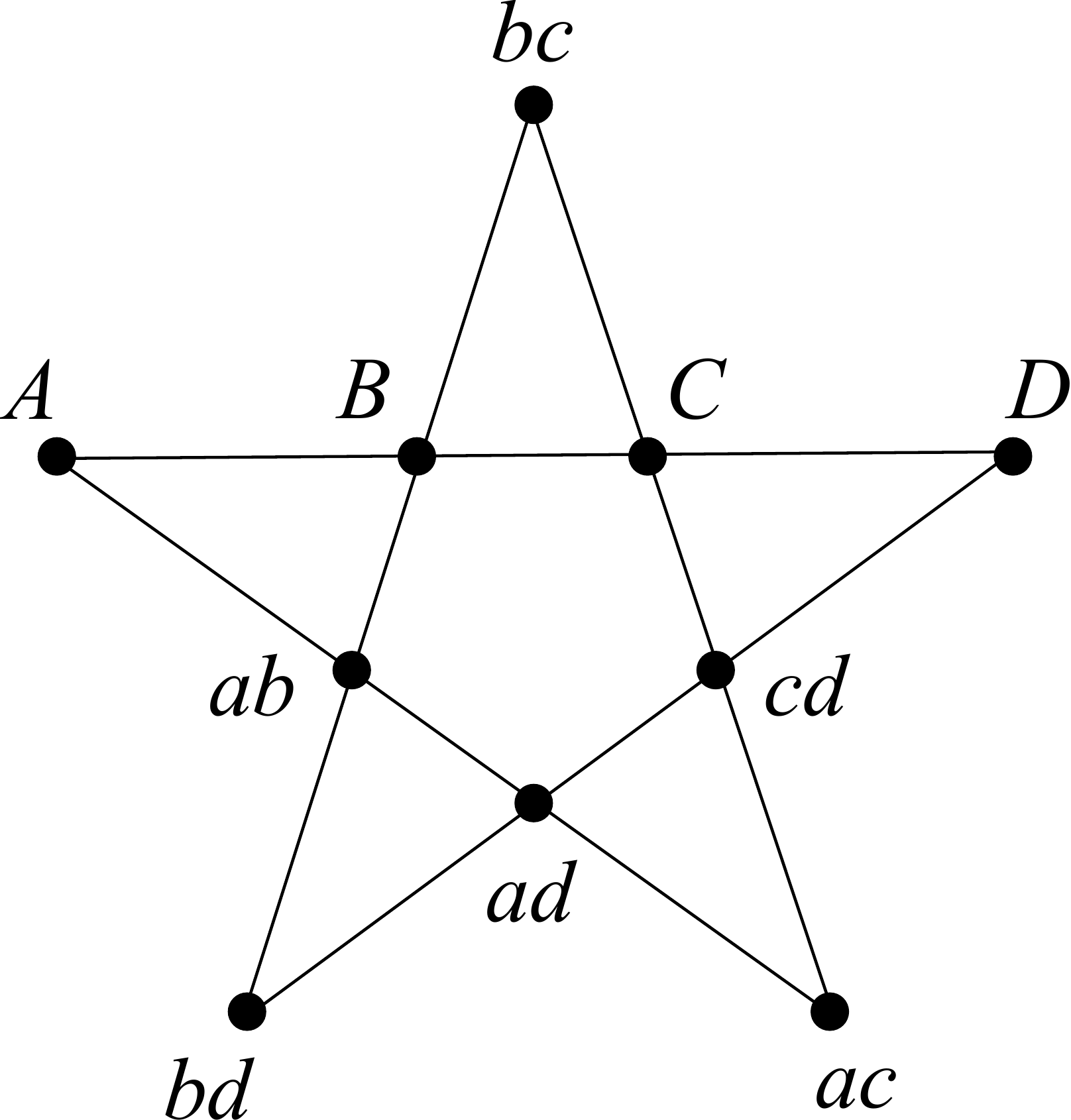}
\caption{Pentagram used in the game to demonstrate that games with PT strategies do not necessarily define tight Bell inequalities.}
\label{fig:pentagram}
\end{figure}

{\em Proof.} The proof uses Theorem \ref{th1} and a nonlocal game with a PT strategy in the $(5,8;5,8)$ Bell scenario. Consider the pentagram in Fig.~\ref{fig:pentagram}. It has five edges and $10$ vertices; four vertices in each edge. In each round of the game $G$, Alice and Bob are asked to output $1$ or $-1$ to each of the four vertices of one edge (not necessarily the same edge). That is, each party must output four bits. The conditions to win $G$ are the following: (I)~The product of the four outputs must be $1$, except when the edge is $\{A,B,C,D\}$. In this case, the product must be $-1$.
(II)~If the parties are asked different edges, both parties must output the same value for the vertex at the intersection of the edges. (III)~If the parties are asked the same edge, Alice's four outputs must be the same as Bob's respective outputs. 


The classical winning probability is $\omega_{\mathrm{C}}(G) = \frac{23}{25}$. However, the quantum correlation $p$, produced with two eight-dimensional systems in the state $|\psi\rangle = \frac{1}{2 \sqrt{2}} \sum_{i=0}^{7} |ii\rangle$, and measuring, on each eight-dimensional system, $A=X \otimes Z \otimes Z$, $B=Z \otimes X \otimes Z$, $C=Z \otimes Z \otimes X$, $D=X \otimes X \otimes X$, $ab=I \otimes I \otimes Z$, $ac=I \otimes Z \otimes I$, $ad=X \otimes I \otimes I$, $bc=Z \otimes I \otimes I$, $bd=I \otimes X \otimes I$, and $cd=I \otimes I \otimes X$, where $X$ and $Z$ are the corresponding Pauli matrices, and $I$ is the identity \cite{Mermin:1990PRLa}, gives $\omega^{(p)}(G) = 1$ and thus provides a PT strategy. 

However, the Bell inequality defined by the game $G$ is not tight.
To show it, let us denote the inputs as follows: $0:\{A,B,C,D\}$, 
$1:\{A,ab,ac,ad\}$, $2:\{ab,B,bc,bd\}$, $3:\{ac,bc,C,cd\}$, and $4:\{ad,bd,cd,D\}$.
Then, the Bell inequality associated to $G$ is
\begin{equation}
\begin{split}
\label{pent}
I_G = &p(A=A|0,1) + p(A=A|1,0) + \ldots + p(cd=cd|4,3) \\
&+ \sum_{x=0}^{4} p(a=b|x=y) \le 23,
\end{split}
\end{equation} 
where, e.g., $p(A=A|0,1)$ is the probability that Alice's and Bob's output for vertex $A$ are equal when Alice's input is $0$ and Bob's input is $1$. $p(a=b|x=y)$ is the probability that Alice's and Bob's outputs are equal, one by one, when Alice's and Bob's inputs are the same.
Inequality \eqref{pent} is saturated by $628$ local vertices, which span a subspace of dimension $460$. However, the dimension of the NS space of the $(5,8;5,8)$ Bell scenario is $1295$. Therefore, inequality \eqref{pent} is not tight. \hfill \qedsymbol

The counterexample used in the proof is not an isolated case. In Appendix~D, we present a general method for non-facet-preserving (i.e., non-tight-preserving) lifting Bell inequalities in which it holds the property that, if the original Bell inequality corresponds to a game with a quantum PT strategy, then the lifted Bell inequality still corresponds to a game with a quantum PT strategy.


{\em Where does FNS=FN=AVN=PT occur?} Combining Theorem~\ref{th1} with previous results \cite{CHTW04,Brassard:2005,Gisin:2007IJQI}, we can conclude that QM does not allow for bipartite FNS=FN=AVN=PT with qubits (a qutrit-qutrit is the smallest quantum system needed) \cite{Brassard:2005}, or if one of the parties has only two settings \cite{Gisin:2007IJQI}, or if all measurements have two outcomes \cite{CHTW04}. However, QM predicts correlations that are arbitrarily close to FNS=FN=AVN=PT in $(m, 2; m, 2)$ using a qubit-qubit maximally entangled state when $m$ tends to infinity \cite{Barrett:2006PRL}. The question is: When does QM allow for bipartite FNS=FN=AVN=PT with a {\em finite} number of settings? The simplest example of quantum bipartite FNS=FN=AVN=PT known occurs in $(3,4;3,4)$ \cite{Cabello:2001PRLa,Cabello:2001PRLb,CinelliPRL2005,YangPRL2005,Aolita:2012PRA,Xu:2022PRL}. But, is there any simpler example?

Gisin, M\'ethot, and Scarani~\cite{Gisin:2007IJQI} made the observation that no result excludes the possibility of quantum PT in the $(3,3;3,2)$ Bell scenario. This led them to raise the question of whether PT can happen in $(3,3;3,2)$. In the light of the equivalences established by Theorem~\ref{th1}, this becomes important in several fields and in relation to several problems. Our third result answers a more general version of this question.


\begin{theorem}
\label{th3}
Quantum mechanics does not allow for FNS=FN=AVN=PT neither in the $(3,3;3,2)$ nor in the $(3,2;3,4)$ Bell scenarios. 
\end{theorem}


{\em Proof:} The proof is based on Theorem~\ref{th1} and in the observation that AVN proofs require correlations whose table of zeros cannot be realized by LHV models. The idea is to identify all tables of zeros that cannot be realized classically unless one of the zeros is removed and then check whether these tables can be realized with a quantum correlation. We will refer to one of such tables as a critical nonlocal table of zeros (CNTZ).
For example, a CNTZ in the $(3,2;3,3)$ Bell scenario is the following:
\begin{center}
\begin{tabular}{c|c||ccc|ccc|ccc}
& $y$ & \multicolumn{3}{c|}{$0$} & \multicolumn{3}{c|}{$1$} & \multicolumn{3}{c}{$2$} \\
\hline
$x$ & \diagbox[innerleftsep=0.3cm,,innerrightsep=0.1cm,innerrightsep=0.2cm,innerwidth=.5cm,height=0.7cm]{$a$}{$b$} & $0$ & $1$ & $2$ & $0$ & $1$ & $2$ & $0$ & $1$ & $2$ \\
\hline \hline
\multirow{2}{*}{$0$} & $0$ & $0$ & $0$ & $\_$ & $\_$ & $\_$ & $\_$ & $\_$ & $\_$ & $\_$ \\
& $1$ & $\_$ & $\_$ & $\_$ & $\_$ & $\_$ & $0$ & $0$ & $\_$ & $\_$ \\
\hline
\multirow{2}{*}{$1$} & $0$ & $\_$ & $\_$ & $0$ & $\_$ & $\_$ & $\_$ & $\_$ & $\_$ & $\_$ \\
& $1$ & $\_$ & $\_$ & $\_$ & $\_$ & $\_$ & $0$ & $0$ & $\_$ & $\_$ \\
\hline
\multirow{2}{*}{$2$} & $0$ & $\_$ & $\_$ & $\_$ & $\_$ & $\_$ & $\_$ & $\_$ & $0$ & $0$ \\
& $1$ & $\_$ & $\_$ & $\_$ & $0$ & $0$ & $\_$ & $\_$ & $\_$ & $\_$ \\
\end{tabular}
\end{center}
Note that it is impossible to
find $f: \{(a |x)\}_{x \in X, a \in A} \cup S_B = \{(b |y)\}_{y \in Y, b \in B} \rightarrow \{0,1\}$ satisfying $\sum_{a} f(a|x)=1$, $\forall x \in X$, and $\sum_{b} f(b|y)=1$, $\forall y \in Y$, without having a pair $\{(a|x),(b|y)\}$ for which $f(a|x)=f(b|y)=1$ and $p(a,b|x,y)=0$.

We wrote a Matlab program that produces all CNTZs, modulo relabelings of inputs, outputs, and parties. The version for $(3,3;3,2)$ is in Appendix~E. We run this program on a high-performance computer and obtained $223$ nonequivalent CNTZs for the $(3,3;3,2)$ Bell scenario. 

To check the quantum realizability of each CNTZ, we used the NPA hierarchy at level 1 (or 2). We found that none of the CNTZs yielded a feasible solution to the corresponding semidefinite programming (SDP) problem in the NPA hierarchy.

Our program for producing all CNTZs relies on the structure of the Bell symmetric group (see Appendix~F). The size of the Bell symmetric group grows rapidly with the addition of more parties, inputs, or outputs. Consequently, the program becomes computationally too demanding for applying it to the $(3,3;3,3)$ and $(3,4;3,3)$ Bell scenarios. Nevertheless, for $(3,2;3,4)$, we still can handle it by using, at a certain step of the program, a sub-group $S'$ of the symmetric group $S$. This provides a faster convergence and results of a manageable size for the subsequent checking of the quantum realizability. See Appendix~E for details. Again, none of the CNTZs in $(3,2;3,4)$ were found to have a quantum realization. \hfill \qedsymbol


Theorem \ref{th3} implies that in $(3, 3; 3, 2)$ and $(3, 2; 3, 4)$ there is a finite gap between the quantum set and the faces of the NS polytope that do not contain local points. This leads to the question of what is the maximum nonlocal content that can be achieved in these scenarios.
We have partially answered this question for $(3,3;3,2)$. The local set for the $(3,3;3,2)$ is fully described by a set of $25$ classes of tight Bell inequalities~\cite{Colbeck:2019PRA,Zambrini:2022XXX}, the facets of the corresponding local polytope. Since we have the half-space representation of the local polytope (i.e., we have the local polytope defined as an intersection of a finite number of half-spaces), we can calculate, for every facet, the corresponding quantum bound (or an upper bound of it) using the Navascu\'es-Pironio-Ac\'{\i}n (NPA) hierarchy \cite{NPA_NJP}. First, we have confirmed that for every facet, the quantum bound is strictly smaller than the NS bound. In addition, we have computed the local, quantum, and NS bounds for each facet (see Appendix~G). The maximum nonlocal content allowed by QM in $(3,3;3,2)$ maximally violating a tight Bell inequality is $q_{\mathrm{NL}}^{\mathrm{min}}=0.598$ (see Appendix~G).

Theorem \ref{th3} leads to the question of whether quantum FNS=FN=AVN=PT is possible in $(3,3;3,3)$. Because of the exponential complexity of the polytopes, the tools used in the proof of Theorem \ref{th3} are not enough for answering this question. Still, we have computed the NS and quantum bounds for $4801183$ classes of local facets in the $(3,3;3,3)$ Bell scenario and found no example of FNS=FN=AVN=PT (see Appendix~H). The lower bound on the number of classes was computed using the tally, i.e., the frequency of distinct coefficients in each inequality, from a total list of $8269146$ facets. The quantum bounds were obtained at level $1+AB$ in the NPA hierarchy.

The method used to prove Theorem \ref{th3} is a useful tool in itself.
The method can be expected to efficiently produce novel quantum FNS=FN=AVN=PT correlations. Why is this important? On the one hand, for the reasons mentioned in (i), (ii), \ldots, (VI') in the introduction. On the other hand, it is also important for graph theory. The graph of exclusivity \cite{CSWPRL2014} of any quantum FNS=FN=AVN=PT correlation has the property that its independence number is strictly smaller than its Lov\'asz and fractional packing numbers, which are equal \cite{Amselem:2012PRL}. Only a few graphs are known with these properties \cite{Amselem:2012PRL}. Our method provides a systematic way to identify new examples. In addition, the method can help to solve another problem \cite{Mancinska} that, after the FNS=FN=AVN=PT equivalence, turns out to have fundamental interest in many fields: Is FNS=FN=AVN=PT possible with nonmaximally entangled states?


{\em Conclusions.} We have shown that four different resources that are crucial for a wide range of applications and results in quantum information and quantum computing are actually equivalent. The term ``resource'' is appropriate as FNS=FN=AVN=PT can be quantified (e.g., with respect to the number of local settings that can be removed while preserving FNS=FN=AVN=PT or by the number of local settings required), used, and consumed. The FNS=FN=AVN=PT equivalence provides a unified perspective about problems in several fields and allows us to combine different tools to investigate this extreme form of Bell nonlocality. 

We have also shown that not all quantum FNS=FN=AVN=PT correlations define tight Bell inequalities. This solves an open problem and shows that finding all quantum FNS=FN=AVN=PT correlations is not easy even if one has the complete description of the local set. We have also solved another open question and demonstrated that QM does not allow for FNS=FN=AVN=PT in $(3, 3; 3, 2)$ and $(3, 2; 3, 4)$. 

In addition, for proving the previous results, we have introduced an efficient method for finding quantum FNS=FN=AVN=PT correlations that is useful in different fields and may help to solve several open problems. Whether FNS=FN=AVN=PT is possible in $(3,3;3,3)$, $(3,3;4,2)$, or $(3,4;3,3)$ remains an open problem.\\




We thank useful conversations with Mateus Ara\'ujo. This work was supported by the EU-funded project \href{10.3030/101070558}{FoQaCiA} and the \href{10.13039/501100011033}{MCINN/AEI} (Project No.\ PID2020-113738GB-I00). E.Z.C.\ acknowledges funding by FCT/MCTES - Fundação para a Ciência e a Tecnologia (Portugal) - through national funds and when applicable cofunding by EU funds under the Project No.\ UIDB/50008/2020, and FCT through Project No.\ 2021.03707.CEECIND/CP1653/CT0002. J.R.G.U.\ acknowledges support from Akademikernas a-kassa (Sweden). L.Y., H.Y.C.\ and R.R.\ acknowledge support from the Research Grants Council, University Grants Committee, Hong Kong. Some computations in this work were performed using research computing facilities offered by Information Technology Services, the University of Hong Kong. A.C.\ acknowledges support from the Wallenberg Center for Quantum Technology (WACQT).


\appendix 


\section{AVN as a nonlocal table of zeros}


Mermin introduced the term ``all versus nothing'' (AVN) \cite{Mermin:1990PRLa} to designate a particular type of proofs of Bell's theorem \cite{Bell:1964PHY}. The term AVN has been in use since the 2000s (see, e.g., \cite{Cabello:2001PRLb,ChenPRL2003,CinelliPRL2005,YangPRL2005}). AVN refers to a particular way of proving the impossibility of Einstein-Podolsky-Rosen (EPR) ``elements of reality'' \cite{Einstein:1935PR}. In fact, Mermin's exact words to describe the proof were ``all versus nothing demolition of EPR'' \cite{Mermin:1990PRLa}. For our purposes, it is important to remember that an EPR element of reality corresponds to an outcome of a local measurement that can be predicted {\em with certainty} from the outcomes of spacelike separated measurements \cite{Einstein:1935PR}.

In this Appendix, we explain the relationship between the characterization of an AVN proof used in the main text [in which an AVN proof is represented by a table of zeros that is not realizable by a local hidden-variable (LHV) model] and the meaning of AVN. We also explain why Hardy's \cite{HardyPRL1993} is not an AVN proof.

An AVN proof consists of $N$ quantum predictions {\em with certainty} with the following property: when $N-1$ (or, more generally, a strict subset) of them are considered, the notion of EPR elements of reality (or the assumption of LHV models) implies a specific prediction with certainty about the $N$th situation. However, this prediction contradicts the corresponding quantum prediction. Therefore, {\em if the $N-1$ predictions are assumed to be valid (since they agree with QM and LHV models)}, then a single experiment testing the $N$th prediction must confirm or exclude EPR elements of reality. Hence the name ``all or nothing.'' Mermin writes: ``The refutation is not only stronger --it is no longer statistical and can be accomplished in a single run'' \cite{Mermin:1990PRLa}.

Now let us show that this meaning is exactly captured by the representation by a table of zeros that is not realizable by any LHV model. We refer to such a table as a {\em nonlocal table of zeros}.

A {\em table of zeros} for the $(|X|, |A|; |Y|, |B|)$ Bell scenario is a matrix with $|X| \times |A|$ rows and $|Y| \times |B|$ columns containing either zeros or empty entries. A zero in the entry $(a,b|x,y)$ indicates that the probability of $(a,b|x,y)$ is zero. 

Given $S=S_A \cup S_B$, with $S_A = \{(a |x)\}_{x \in X, a \in A}$ and $S_B = \{(b |y)\}_{y \in Y, b \in B}$, a table of zeros is not realizable by an LHV model if, for every assignment $f: S \rightarrow \{0,1\}$ satisfying $\sum_{a} f(a|x)=1$, $\forall x \in X$, and $\sum_{b} f(b|y)=1$, $\forall y \in Y$, there is a pair $\{(a|x),(b|y)\}$ for which $f(a|x)=f(b|y)=1$ and $p(a,b|x,y)=0$.


For example, the quantum correlations leading to Hardy's proof of Bell nonlocality \cite{HardyPRL1993} produce the following table of zeros:
\begin{center}
\begin{tabular}{c|c||cc|cc}
& $y$ & \multicolumn{2}{c|}{$0$} & \multicolumn{2}{c}{$1$} \\
\hline
$x$ & \diagbox[innerleftsep=0.3cm,innerrightsep=0.1cm,innerrightsep=0.2cm,innerwidth=.5cm,height=0.7cm]{$a$}{$b$} & $0$ & $1$ & $0$ & $1$ \\
\hline \hline
\multirow{2}{*}{$0$} & $0$ & $\_$ & $\_$ & $\_$ & $0$ \\
& $1$ & $\_$ & $\_$ & $\_$ & $\_$ \\
\hline
\multirow{2}{*}{$1$} & $0$ & $\_$ & $\_$ & $0$ & $\_$ \\
& $1$ & $0$ & $\_$ & $\_$ & $\_$ \\
\end{tabular}
\end{center}
However, Hardy's proof is not an AVN proof because {\em there are local models realizing the table of zeros}. For example, if we use {\color{green}green} to indicate the outcome determined by the LHV model, then the following local assignment realizes all zeros, since it assigns one outcome to each of the four local measurements and there is no zero in the intersection of any two greens:
\begin{center}
\begin{tabular}{c|c||cc|cc}
& $y$ & \multicolumn{2}{c|}{$0$} & \multicolumn{2}{c}{$1$} \\
\hline
$x$ & \diagbox[innerleftsep=0.3cm,,innerrightsep=0.1cm,innerrightsep=0.2cm,innerwidth=.5cm,height=0.7cm]{$a$}{$b$} & {\color{green} $0$} & $1$ & $0$ & {\color{green} $1$} \\
\hline \hline
\multirow{2}{*}{$0$} & $0$ & $\_$ & $\_$ & $\_$ & $0$ \\
& {\color{green} $1$} & $\_$ & $\_$ & $\_$ & $\_$ \\
\hline
\multirow{2}{*}{$1$} & {\color{green} $0$} & $\_$ & $\_$ & $0$ & $\_$ \\
& $1$ & $0$ & $\_$ & $\_$ & $\_$ \\
\end{tabular}
\end{center}

In contrast to that, in the simplest known quantum bipartite AVN proof, which occurs in the $(3,4;3,4)$ Bell scenario, the table of zeros is
\begin{center}
\begin{tabular}{c|c||cccc|cccc|cccc}
& $y$ & \multicolumn{4}{c|}{$0$} & \multicolumn{4}{c|}{$1$} & \multicolumn{4}{c}{$2$} \\
\hline
$x$ & \diagbox[innerleftsep=0.3cm,,innerrightsep=0.1cm,innerrightsep=0.2cm,innerwidth=.5cm,height=0.7cm]{$a$}{$b$} & $0$ & $1$ & $2$ & $3$ & $0$ & $1$ & $2$ & $3$ & $0$ & $1$ & $2$ & $3$ \\
\hline \hline
\multirow{4}{*}{$0$} & $0$ & $\_$ & $\_$ & $0$ & $0$ & $\_$ & $0$ & $\_$ & $0$ & $\_$ & $\_$ & $0$ & $0$ \\
& $1$ & $\_$ & $\_$ & $0$ & $0$ & $0$ & $\_$ & $0$ & $\_$ & $0$ & $0$ & $\_$ & $\_$ \\
& $2$ & $0$ & $0$ & $\_$ & $\_$ & $\_$ & $0$ & $\_$ & $0$ & $0$ & $0$ & $\_$ & $\_$ \\
& $3$ & $0$ & $0$ & $\_$ & $\_$ & $0$ & $\_$ & $0$ & $\_$ & $\_$ & $\_$ & $0$ & $0$ \\
\hline
\multirow{4}{*}{$1$} & $0$ & $\_$ & $0$ & $\_$ & $0$ & $\_$ & $\_$ & $0$ & $0$ & $\_$ & $0$ & $\_$ & $0$ \\
& $1$ & $0$ & $\_$ & $0$ & $\_$ & $\_$ & $\_$ & $0$ & $0$ & $0$ & $\_$ & $0$ & $\_$ \\
& $2$ & $\_$ & $0$ & $\_$ & $0$ & $0$ & $0$ & $\_$ & $\_$ & $0$ & $\_$ & $0$ & $\_$ \\
& $3$ & $0$ & $\_$ & $0$ & $\_$ & $0$ & $0$ & $\_$ & $\_$ & $\_$ & $0$ & $\_$ & $0$ \\
\hline
\multirow{4}{*}{$2$} & $0$ & $\_$ & $0$ & $0$ & $\_$ & $\_$ & $0$ & $0$ & $\_$ & $0$ & $\_$ & $\_$ & $0$ \\
& $1$ & $\_$ & $0$ & $0$ & $\_$ & $0$ & $\_$ & $\_$ & $0$ & $\_$ & $0$ & $0$ & $\_$ \\
& $2$ & $0$ & $\_$ & $\_$ & $0$ & $\_$ & $0$ & $0$ & $\_$ & $\_$ & $0$ & $0$ & $\_$ \\
& $3$ & $0$ & $\_$ & $\_$ & $0$ & $0$ & $\_$ & $\_$ & $0$ & $0$ & $\_$ & $\_$ & $0$ \\
\end{tabular}
\end{center}
and no LHV model can reproduce it. The fact that, if we remove any of the predictions (the $4 \time 4$ squares inside the table), then there is an LHV model for the resulting table of zeros implies that, for any remaining set of predictions with certainty, there is a local model.


\section{Equivalence between FNS and FN}


FNS are the nonlocal correlations that live in a face of the NS polytope that does not contain local points. A correlation $p$ is FN if and only if its nonlocal content $q_{\mathrm{NL}}^{\mathrm{min}}$ is $1$, where $q_{\mathrm{NL}}^{\mathrm{min}}:=1-q_{\mathrm{L}}^{\mathrm{max}}$ and $q_{\mathrm{L}}^{\mathrm{max}}$ is the maximum local weigh over all decompositions of the form
\begin{equation}\label{decompose}
p(a,b|x,y)=q_{\mathrm{L}} p_{\mathrm{L}}(a,b|x,y)+(1-q_{\mathrm{L}})p_{\mathrm{NL}}(a,b|x,y).
\end{equation}
$p_{\mathrm{L}}(a,b|x,y)$ represents local correlations and $p_{\mathrm{NL}}(a,b|x,y)$ represents nonlocal NS correlations.

{\em $p$ is FN $\Rightarrow$ $p$ is FNS.} If $p$ is FN, then $p$ must lie in a face $\mathcal{F}$ of the NS polytope. This is because the NS polytope is defined by the conditions of non-negativeness and normalization of the probabilities, and the conditions of NS. The normalization and NS conditions are equalities and the non-negative conditions are inequalities. 
Suppose, on the contrary, that $p$ lies in the interior of the NS polytope, then none of the non-negative conditions are saturated, i.e., $p(a,b|x,y)>0,\forall a,b,x,y$. Denote the minimum entry of $p$ by $p_{\mathrm{min}}$, $p$ can be decomposed as
\begin{equation}
\begin{split}
p(a,b|x,y)=&p_{\mathrm{min}}\cdot |A||B| \frac{p_{\mathbb{I}}(a,b|x,y)}{|A||B|}+\\
&(1-p_{\mathrm{min}}\cdot |A||B|)\frac{p(a,b|x,y)-p_{\mathrm{min}}p_{\mathbb{I}}(a,b|x,y)}{1-p_{\mathrm{min}}\cdot |A||B|},
\end{split}
\end{equation}
where $\frac{p_{\mathbb{I}}(a,b|x,y)}{|A||B|}=\frac{1}{|A||B|}\forall a,b,x,y$. It is clear that both $\frac{p_{\mathbb{I}}(a,b|x,y)}{|A||B|}$ and $\frac{p(a,b|x,y)-p_{\mathrm{min}}p_{\mathbb{I}}(a,b|x,y)}{1-p_{\mathrm{min}}\cdot |A||B|}$ are NS correlations, which satisfy the non-negative conditions, normalization conditions, and NS conditions. On the other hand, $\frac{p_{\mathbb{I}}(a,b|x,y)}{|A||B|}$ is a local correlation. Therefore, the local content of $p$ is at least $p_{\mathrm{min}}|A||B|$. In this way, $p$ is not FN.

Secondly, there are two possibilities that $p$ lies in a face $\mathcal{F}$ of the NS polytope: (i) $\mathcal{F}$ contains a local point or (ii) $\mathcal{F}$ does not contain a local point. In case (i), one can define a face $\mathcal{F}'$ as the convex hull of all extreme nonlocal points of $\mathcal{F}$. There must be $p\in \mathcal{F}$, otherwise, there would exist a convex decomposition of $p$ that contains a local extreme point of $\mathcal{F}$, which is impossible by the definition of FN. So $p$ is FNS. For case (ii), we directly obtain that $p$ is FNS.

{\em $p$ is FNS $\Rightarrow$ $p$ is FN.} Suppose that $p$ is in a face of the NS polytope that does not contain local points. Let us denote this face by $\mathcal{F}$. Then, for any decomposition of $p$ of the form Eq.~\eqref{decompose}, any correlation on the right-hand-side of Eq.~\eqref{decompose} belongs to $\mathcal{F}$. As $\mathcal{F}$ contains only nonlocal correlations, $q_{\mathrm{L}}=0$ for all decomposition of $p$ of the form Eq.~\eqref{decompose}. Then, $q_{\mathrm{L}}^{\mathrm{max}}=0$, $p$ is FN.


\section{Equivalence between AVN and PT}


An AVN proof is represented by a table of zeros that is not realizable in LHV model but realizable in QM. The table is a matrix with $|X|\times |A|$ rows and $|Y|\times |B|$ columns containing either zeros or nondefined entries. Let us denote this matrix by $M$ and its entries by $M(xa,yb)$. Then, $M(xa,yb)\in\{0,\_ \}$, where $\_$ represents the undefined entries.

A game $G$ allows for a PT strategy if $\omega_{\mathrm{C}}(G)<\omega_{\mathrm{Q}}(G)=1$.

{\em $p$ allows for an AVN proof $\Rightarrow$ $p$ allows for a PT strategy.} Let us define a set $S_M$ that includes all zero entries of $M$, i.e., $S_{M}:=\{(a,b,x,y):M(xa,yb)=0\}$. From the AVN proof, we know that there exists a quantum correlation $p_{\mathrm{Q}}$ that $p_{\mathrm{Q}}(a,b|x,y)=0,\forall (a,b,x,y)\in S_M$ and for all LHV correlation $p_{\mathrm{L}}$ the inequality $\sum_{(a,b,x,y)\in S_M}p_{\mathrm{L}}(a,b|x,y)>0$ holds.
We can define a game $G=(X\times Y, A\times B,\pi, W)$ whose input and output set is the same as the AVN proof, the input distribution is $\pi(x,y)=\frac{1}{|X||Y|},\forall x,y$ and whose winning conditions are 
\begin{equation}
W(a,b,x,y)=\begin{cases}
1 & \text{if } (a,b,x,y)\notin S_M,\\
0 &\text{if } (a,b,x,y)\in S_M.\\
\end{cases}
\end{equation}
The quantum winning probability by using the quantum correlation $p_{\mathrm{Q}}$ is 
\begin{equation}
\begin{split}
\omega_{\mathrm{Q}}(G)&=\sum_{x,y,a,b}\frac{1}{|X||Y|}p_{\mathrm{Q}}(a,b|x,y)W(a,b,x,y)\\
&=\sum_{(a,b,x,y)\notin S_M}\frac{1}{|X||Y|}p_{\mathrm{Q}}(a,b|x,y)\\
&=\frac{1}{|X||Y|} \sum_{x,y}1=1.\\
\end{split}
\end{equation}
The second equality holds because of the definition of winning conditions. The third equality holds because of the normalization conditions and that $p_{\mathrm{Q}}(a,b|x,y)=0,\forall (a,b,x,y)\in S_M$. On the other hand, the classical winning probability is 
\begin{equation}
\begin{split}
\omega_{\mathrm{C}}(G)&=\max_{p_{\mathrm{L}}} \sum_{x,y,a,b}\frac{1}{|X||Y|}p_{\mathrm{L}}(a,b|x,y)W(a,b,x,y)\\
&=\max_{p_{\mathrm{L}}} \sum_{(a,b,x,y)\notin S_M}\frac{1}{|X||Y|}p_{\mathrm{L}}(a,b|x,y)\\
&=\max_{p_{\mathrm{L}}} \left[1-\sum_{(a,b,x,y)\in S_M}\frac{1}{|X||Y|}p_{\mathrm{L}}(a,b|x,y)\right]\\
&=1-\frac{1}{|X||Y|} \max_{p_{\mathrm{L}}}\sum_{(a,b,x,y)\in S_M}p_{\mathrm{L}}(a,b|x,y)<1.
\end{split}
\end{equation}
The last inequality holds because $\sum_{(a,b,x,y)\in S_M}p_{\mathrm{L}}(a,b|x,y)>0,\forall p_{\mathrm{L}}$. Therefore, we conclude that $G$ is a game that has a PT strategy.

{\em $p$ allows for a PT strategy $\Rightarrow$ $p$ allows for an AVN proof.} From the game $G$ having a PT strategy we can define the following matrix $M$ with $|X|\times|A|$ rows and $|Y|\times |B|$: 
\begin{equation}
M(xa,yb)=\begin{cases}
0 & \text{if }W(a,b,x,y)=0 \wedge \pi(x,y)>0,\\
\_ & \text{otherwise.}
\end{cases}
\end{equation} 
Then, we can define the set $S_{M}:=\{(a,b,x,y):M(xa,yb)=0\}$, which includes all zero entries of $M$. 

Since $\omega_{\mathrm{Q}}(G)=1$, there exists one quantum correlation $p_{\mathrm{Q}}$ such that $p_{\mathrm{Q}}(a,b|x,y)=0,\forall (a,b,x,y)\in S_M$.

On the other hand, since $\omega_{\mathrm{C}}(G)<1$, we have 
\begin{equation}
\begin{split}
\omega_{\mathrm{C}}(G)=&\max_{p_{\mathrm{L}}} \sum_{x,y,a,b}\pi(x,y)p_{\mathrm{L}}(a,b|x,y)W(a,b,x,y)\\
=&\max_{p_{\mathrm{L}}} \sum_{\substack{x,y,a,b \text{ that } \\
W(a,b,x,y)\neq 0\wedge \pi(x,y)>0}}\pi(x,y)p_{\mathrm{L}}(a,b|x,y)\\
=&1-\max_{p_{\mathrm{L}}} \sum_{\substack{x,y,a,b \text{ that } \\
W(a,b,x,y)= 0 \wedge \pi(x,y)>0}}\pi(x,y)p_{\mathrm{L}}(a,b|x,y)<1\\
\Rightarrow& \max_{p_{\mathrm{L}}} \sum_{(a,b,x,y)\in S_M} \pi(x,y)p_{\mathrm{L}}(a,b|x,y)>0\\
\Rightarrow& \sum_{(a,b,x,y)\in S_M} p_{\mathrm{L}}(a,b|x,y)>0,\forall p_{\mathrm{L}}.
\end{split}
\end{equation}
Therefore, $M$ represents an AVN proof. 


\section{Non-facet preserving lifting of Bell inequalities}
\label{lifting}


In \cite{pironio2005lifting}, Pironio introduced a method for ``lifting'' a Bell inequality in a given Bell scenario into a Bell inequality in a Bell scenario with more observers, measurement settings, or measurement outcomes. Pironio's method has the property that, if the original Bell inequality was tight (a facet of the local polytope), then the lifted Bell inequality is also tight. Here, we introduce a lifting method that does {\em not} preserve tightness and apply it to a tight Bell inequality associated to a game having a quantum PT strategy in order to produce a a non-tight Bell inequality associated to a game having a quantum PT strategy.

As described in the main text, any bipartite nonlocal game $G$ can be written as a 4-tuple $G=(X\times Y,A\times B,\pi,W)$, where $X$ and $Y$ are the inputs sets for space-like separated players (Alice and Bob), $A$ and $B$ are the corresponding outputs sets, $\pi: \pi\left(X,Y\right)$ is the input distribution (here we simply take it always to be uniform in the support of inputs) and $W: W\left(A,B,X,Y\right)\in\{0,1\}$ is the winning condition function. The associated Bell inequality is characterized by the winning probability of the game $\omega(G)=\sum_{a\in A,b\in B,x\in X, y\in Y}\pi\left(x,y\right) W(a,b,x,y) p(a,b|x,y)$.

Given a game $G$, let us take two copies of it, called $G_1$ and $G_2$, and define a new game $\widetilde{G}$ as follows: $\widetilde{G}=(\widetilde{X}\times\widetilde{Y}, A\times B,\widetilde{\pi},\widetilde{W})$, where $\widetilde{X}=X_1\cup X_2$, $\widetilde{Y}=Y_1\cup Y_2$, $\widetilde{\pi}$ is the uniform distribution over the inputs, and $\widetilde{W}(a,b,x,y)$ is
\begin{equation}
\begin{cases} W(a,b,x,y)& \text{if\,} (x,y)\in(X_1,Y_1) \ \text{or}\ (x,y)\in(X_2,Y_2), \\ 1 & \text{otherwise}. \\ \end{cases} 
\end{equation} 
Then, the best local winning strategies of $\widetilde{G}$ are the optimal ones for winning $G_1$ and $G_2$ separately. That is, if there are $n_{LHV}$ local deterministic boxes that give the maximal local value of $\omega({G})$, there are $n_{LHV}^2$ local deterministic boxes that maximally win $\widetilde{G}$. The optimal quantum strategies for $\widetilde{G}$ will be the optimal quantum strategies for $G_1$ and $G_2$ separately.

The dimension of the local polytope in the $(m,k;m,k)$ Bell scenario is \cite{Collins:2004JPA}
\begin{equation}
d_{m,k}=\left[m\left(k-1\right)+1\right]^2-1.
\end{equation}
Suppose that the Bell inequality associated to a game $G$ described above, with $|X|=|Y|=m$ and $|X|=|Y|=k$, is tight. Then, all the local deterministic strategies that give the maximal value of $\omega(G)$ compose the hyperplane of dimension $d_{m,k}-1$ of the local polytope $\mathrm{L}(m,k;m,k)$. 
This means that, if one can arrange each local deterministic strategy that gives the maximal value of $\omega(G)$ as a vector and arrange $n_{LHV}$ vectors as rows in a matrix, then the rank of the matrix $M_G$ should be $\rank(M_G)=d_{m,k}$. 

If we similarly arrange each local deterministic strategy that gives the maximal winning value of game $\widetilde{G}$ as a vector, and arrange $n_{LHV}^2$ vectors as rows of matrix $M_{\widetilde{G}}$. The matrix can be written as $M_{\widetilde{G}}=[v_{(1,1),i}\oplus v_{(2,2),j}\oplus v_{(1,2),p}\oplus v_{(2,1),q}]$, where $v_{(1,1),i}$ is the vector corresponds to the input $(X_1,Y_1)$, $v_{(2,2),j}$ corresponds to the input $(X_2,Y_2)$, and $v_{(1,2),p}$ [or $v_{(2,1),q}$] corresponds to the input $(X_1,Y_2)$ [or $(X_2,Y_1)$]. For each row of $M_{\widetilde{G}}$ (i.e., each optimal local deterministic strategy of $\widetilde{G}$), once $v_{(1,1),i}$ and $v_{(2,2),j}$ are given, then the other two $v_{(1,2),p}$ and $v_{(2,1),q}$ are fixed. According to the lifting described above, $v_{(1,1),i}$ and $v_{(2,2),j}$ are rows of the matrix $M_G$ of the original game $G$. More importantly, they are chosen from rows of $M_G$ independently. Since $n$ linearly independent vectors define a hyperplane of dimension $n-1$, then, for the two matrices $A$ and $B$, we have $\rank(A\oplus B)=\rank(A)+\rank(B)-1$ by the Cartesian product of the corresponding hyperplanes. Therefore, if the dimension of the face supported by the optimal local realizations of the game $G$ is 
the dimension of the facet minus $t$, then the dimension of the face for the new game $\widetilde{G}$ is at least the dimension of the facet minus $2t+1$. 

Suppose that the original game $G$ is tight, i.e., $t=0$. Then,
\begin{equation} 
\rank(M_{\widetilde{G}})\leq d_{2m, k}-1=\left[2m\left(k-1\right)+1\right]^2-2.
\end{equation}
Therefore, the lifting is not facet-preserving. 

Interestingly, we can use this lifting strategy recursively, i.e., we can take $n$ copies of the tight game $G_1,G_2,\ldots,G_n$ and define a new game $\widetilde{G}_n=(\widetilde{X}_n\times\widetilde{Y}_n, A\times B,\widetilde{\pi}_n,\widetilde{W}_n)$, where $\widetilde{X}_n=X_1\cup X_2\cup\ldots \cup X_n, \widetilde{Y}_n=Y_1\cup Y_2\cup\ldots \cup Y_n$, $\widetilde{\pi}_n$ is the uniform distribution over the new inputs set, and 
\begin{equation}
\widetilde{W}_n(a,b,x,y)=\begin{cases} W(a,b,x,y)& \text{if} (x,y)\in(X_i,Y_i) \forall i\in[n], \\ 1 & \text{otherwise}. \\ \end{cases}
\end{equation} 
For $\widetilde{G}_n$, there are $n_{LHV}^n$ optimal local deterministic boxes that achieve its optimal value, and the hyperplane defined by them in the $(nm,k;nm,k)$ Bell scenario is upper bounded by $d_{nm,k}-(n-1)$. Clearly, if $G$ is a PT game, then $\widetilde{G}_n$ is also a PT game (the optimal quantum strategy is simply the optimal quantum strategies for $G_i \forall i\in[n]$ separately). 

In the following, we present two examples of how our lifting method is applied. The second example lifts a Bell inequality that has a PT strategy.

(I) The CHSH Bell inequality is the facet inequality in the $(2,2;2,2)$ Bell scenario. The winning condition for the CHSH game is 
\begin{equation}
W(a,b,x,y)=\begin{cases} 1 & \text{if\,} a\oplus b=x\cdot y, \\ 0 & \text{otherwise}. \\ \end{cases}
\end{equation}
There are 8 local deterministic boxes that optimally win this game with $\omega(CHSH)=\frac{3}{4}$ [the inputs are uniformly distributed $\pi\left(x,y\right)=1/4,\forall x,y$], and they form the facet of $\mathrm{L}(2,2;2,2)$, i.e., $\rank(M_{CHSH})=d_2$. If we consider 2, 3, and 4 copies of the CHSH game and lift the game as described above, then $\rank(M_{\widetilde{CHSH}_2})=23=d_{4,2}-1$, $\rank(M_{\widetilde{CHSH}_3})=
46=d_{6,2}-2$, and $\rank(M_{\widetilde{CHSH}_4})=77=d_{8,2}-3$, respectively.

\begin{figure}[H]\label{MS_table}
\centering
\includegraphics[width=0.33\textwidth]{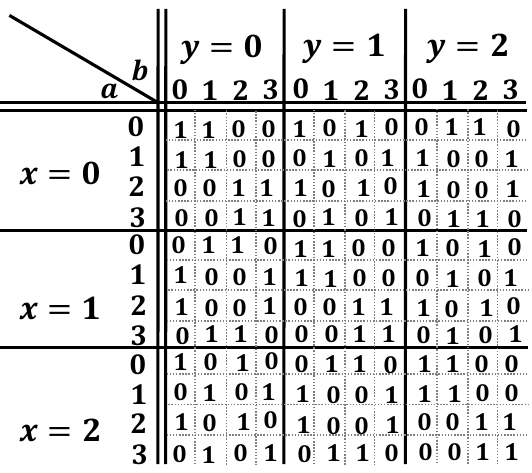}
\caption{Winning condition $W(A,B,X,Y)$ for the magic square game.}
\end{figure}

(II) The Bell inequality associated to the magic square game $MS$ \cite{Cabello:2001PRLb,Aravind:2004AJP} is a tight inequality \cite{Gisin:2007IJQI} in the $(3,4;3,4)$ Bell scenario. The winning condition function $W(A,B,X,Y)$ is in Fig.~\ref{MS_table}. There are $144$ local deterministic strategies that achieve the maximal value of $\omega(MS)=\frac{8}{9}$ [the inputs are uniformly distributed $\pi\left(x,y\right)=1/9,\forall x,y$]. Arrange each local deterministic strategy as a vector and then arrange the $144$ vectors in a matrix. The rank of the matrix is $\rank(M_{MS})=99$, which equals to $d_{3,4}=99$. For the lifted game $\widetilde{MS}_2$, there are $144^2=20736$ local deterministic strategies achieving the optimal winning value and $\rank(M_{\widetilde{MS}_2})=359= d_{6,4}-1$.


\section{Algorithm for enumerating critical nonlocal tables of zeros}
\label{app:c}


To simplify the enumeration of the critical nonlocal tables of zeros (CNTZs) and minimize the computational overhead, Algorithm~\ref{critical_zero_table} partitions the table into two sub-regions, denoted $R\_blue$ and $R\_red$. We initially enumerate the CNTZs within these sub-regions, focusing solely on events occurring within these indicated sub-regions. 

It is important to note that, if a table is a CNTZ, then its respective sub-region must also be a CNTZ if we only consider events restricted to that sub-region. However, the converse is not true: combining CNTZs from the sub-regions does not always yield a CNTZ for the whole. Consequently, after identifying all CNTZs in both $R\_blue$ and $R\_red$, we designate their joint pair as $pretable$ and proceed to construct full-table CNTZs based on these.
On the other hand, the two sub-regions $R\_blue$ and $R\_red$ are symmetric. Therefore, we can relabel Alice's output indices to shift events from $R\_blue$ to $R\_red$ or vice versa. Once we have enumerated all CNTZs within $R\_blue$, we compute all orbits of each table in the symmetric group \(S\_blue\) and subsequently shift them to $R\_red$.

To mitigate the computational burden, at one step (function $\mathrm{GroupReduction}$) of Algorithm~\ref{critical_zero_table}, we use a sub-group $S'$ of the symmetric group $S$ to reducing $ZeroTable$ to $ZeroTable\_sub$ faster. This ensures that $CriticalZeroTable \subseteq ZeroTable\_sub\subseteq ZeroTable$ and that $ZeroTable\_sub$ is of a manageable size for checking of the quantum realizability.


\begin{figure}[htbp]
\centering
\includegraphics[width=0.32\textwidth]{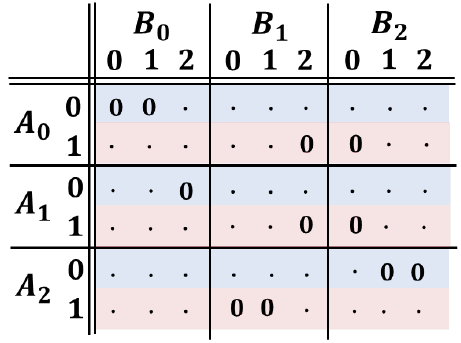}
\caption{An example of CNTZ in the $(3,2;3,3)$ Bell scenario. The two sub-regions $R\_blue$ and $R\_red$ used in Algorithm~\ref{critical_zero_table} are indicated with blue and red backgrounds, respectively.}
\label{region}
\end{figure}


\begin{widetext}
\begin{algorithm}[H]
\SetKwFunction{GenerateZPB}{GenerateZPB}
\SetKwFunction{GroupReduction}{GroupReduction}
\SetKwProg{Fn}{Function}{:}{end}
\SetKwInOut{Input}{Input}
\SetKwInOut{Output}{Output}

\tcp{Step 0: Preparation. }
Label the events in the table. \tcp{A CNTZ is a vector of these labels.}
Denote the full table as region $R$ and split it into two sub-regions $R\_blue$ and $R\_red$ \tcp{Whenever the region is indicated, only events within the indicated region are taken into account.}
Generate the symmetric group $S$ for the whole table and $S\_blue$ for sub-region $R\_blue$\tcp{See Appendix~\ref{group} for details on the symmetric group.}

\tcp{Step 1: Generate CNTZs in the blue region.}
$ZeroTable\_blue:=$\GenerateZPB{$[\ ]$, $R\_blue$}\;
$CriticalZeroTable\_blue:=$\GroupReduction{$ZeroTable\_blue$, $S\_blue$}\;
\tcp{Step 2: Generate pre CNTZs in the full table.}
Initialize $CriticalZeroTable\_red$ as an empty cell set\;
\For{\rm{each element} $table$ \rm{in} $CriticalZeroTable\_blue$}{
$OrbitTable\_blue:=$ orbits of $table$ in the symmetric group $S\_blue$\;
Change the labels of each element $orbittable\_blue$ of $OrbitTable\_blue$ such that the tables are moved from $R\_blue$ to the region $R\_red$, and add the shifted orbit tables to $CriticalZeroTable\_red$\;
}
Initialize $PreTable$ as an empty cell set\;
\For{\rm{each element} $table\_blue$ \rm{in} $CriticalZeroTable\_blue$}{
\For{\rm{each element} $table\_red$ \rm{in} $CriticalZeroTable\_blue$}{
$pretable:=[table\_blue, table\_red]$\;
Add $pretable$ to $PreTable$;
}
}
$PreTable=$\GroupReduction{$PreTable$,$S$}\;
\tcp{Step 3: Generate CNTZs based on pre CNTZs in the full table.}
Initialize $ZeroTable$ as an empty cell set\;
\For{\rm{each element} $pretable$ \rm{in set} $PreTable$}{
Add \GenerateZPB{$pretable$, $R$} to $ZeroTable$\;
}
$CriticalZeroTable:=$\GroupReduction{$ZeroTable$, $S$}\;

\BlankLine
\hrule
\BlankLine
\tcp{Add zeros to the pre CNTZ to make it a CNTZ in the indicated region.}
\setcounter{AlgoLine}{0}
\Fn{$ZeroTable$=\GenerateZPB{$pretable$, $region$}}{
Initialize $ZeroTable$ as an empty cell set\;
$LocalStrategy:=$ all feasible local deterministic tables in the $region$ that satisfy the zero constraints in $pretable$\;
\eIf{$LocalStrategy$ \rm{is empty}}{Add $pretable$ to $ZeroTable$\; }{
$localstrategy:=$Randomly pick one local deterministic table in $LocalStrategy$\;
\For{\rm{each element} $e$ \rm{of} $localstrategy$}{
Add \GenerateZPB{$[pretable, e]$, $region$} to $ZeroTable$\;}
}
}
\BlankLine
\hrule
\BlankLine
\tcp{Use the symmetric group for deleting the duplicated tables.}
\setcounter{AlgoLine}{0}
\Fn{$OutputTable=$\GroupReduction{$InputTable$, $Group$}}{
Sort $InputTable$ by the length of its elements\;
Initialize $OutputTable$ as an empty cell set\;
\While{$InputTable$ \rm{is not empty}}{
$table:=InputTable\{1\}$\;
Add $table$ to $OutputTable$\;
$OrbitTable:=$ orbits of $table$ in the symmetric group $Group$\;
\For{\rm{each element} $orbittable$ \rm{of} $OrbitTable$}{
Filter the element $input$ in $InputTable$ if it takes $orbittable$ as a subset, i.e., $input\supseteq orbittable$\;
} 
}
}
\caption{Enumerate all CNTZs in the $(3,3;3,2)$ Bell scenario}\label{critical_zero_table}
\end{algorithm}
\end{widetext}


\section{The Bell symmetric group}\label{group}


Here, we review the group structure of the Bell symmetric group and the bound of its minimal number of generators.

The Bell symmetric group can be expressed using the wreath product of symmetric groups of different order. Then, let us first recall the definition of wreath product of symmetric groups. Let $S_n$ be the symmetric group on $n$ elements.
\begin{definition}
Let $S\leq S_n$ and $G$ be an arbitrary group. Let us define
\begin{equation}
G \wreath S := G^n \rtimes S
\end{equation}
Consequently, the elements of $G\wreath S$ are of the form $(g_1,\ldots,g_n;\sigma)$, where $g_i\in G$ and $\sigma \in S$. The action of $\sigma \in S$ on $(g_1,\ldots,g_n)\in G^n$ is given by
\begin{equation}
\sigma\cdot (g_1,\ldots,g_n)=(g_{\sigma(1)},\ldots,g_{\sigma(n)}).
\end{equation}
The operation on $G\wreath S$ is given by
\begin{equation}
(g_1,\ldots,g_n;\sigma)(h_1,\ldots,h_n;\tau)=(g_1h_{\sigma(1)},\ldots,g_nh_{\sigma(n)};\sigma\tau).
\end{equation}
The inverse of an element is given by
\begin{equation}
(g_1,\ldots,g_n;\sigma)^{-1}=(g^{-1}_{\sigma^{-1}(1)},\ldots,g^{-1}_{\sigma^{-1}(n)};\sigma^{-1}). 
\end{equation}
\end{definition}

Now consider the $(3,2;3,2)$ Bell scenario. Bell scenarios are defined modulo permutation of settings and outputs. Therefore, if $S$ is the Bell symmetric group of $(3,2;3,2)$, then $S$ includes four types of symmetries: permutations of Alice's inputs, permutations of Bob's inputs, permutations of Alice's outputs, and permutations of Bob's outputs. $S$ does not include the permutation of the parties. Since a permutation in Alice's side commutes with a permutation in Bob's side, then the group structure of $S$ is given by
\begin{equation}
S=(S_2 \wreath S_3) \times (S_3 \wreath S_3),
\end{equation}
where the $S_2 \wreath S_3$ is the group generated by the permutations in Alice's side while $S_3 \wreath S_3$ is the group generated by the permutations in Bob's side. Since a permutation in Alice's side commutes with a permutation in Bob's side, we take direct product between these groups.
For any group $G$, we denote by $d(G)$ the minimal number of generators of it. Using Theorem~1.1 in~\cite{East:2021NZJM}, we have
$d(S_2 \wreath S_3) = d(S_3 \wreath S_3) = 2$.
Therefore, 
\begin{equation}
2 = d(S_2 \wreath S_3) \leq d(S) \leq d(S_2 \wreath S_3)+d(S_3 \wreath S_3) = 4.
\end{equation}


\section{Complete list of facets in $(3,3;3,2)$}
\label{sec:appA}


Table \ref{tab:appA} shows the local (L), quantum (Q), and nonsignaling (NS) bounds for every nontrivial (excluding positivity) class of facet of the (3,3;3,2) Bell scenario. The names of the classes and inequalities follow those in \cite{Zambrini:2022XXX}. Table \ref{tab:appA} also includes two measures of Bell nonlocality for the correlations that attain the maximum violation of these inequalities: the ratio between the quantum and NS bound $(Q/NS)$, and the nonlocal content $q_{\mathrm{NL}}^{\mathrm{min}}$ \cite{Elitzur:1992PLA}. We see that class 18 exhibits the largest quantum to NS bound ratio, $70.73\ \%$. More importantly, none of the optimal quantum strategies attains the optimal NS bound. The nonlocal content presented correspond to the optimal quantum strategies for ququarts, which saturate the NPA bound except in the case of the $I_{3322}$ inequality, i.e., classes 5 to 7. To find the optimal quantum strategies, we employed a seesaw of SDP problems. We used Matlab's CVX \cite{CVX} and the SDPT3 solver \cite{SDPT3}. The nonlocal contents were then calculated for the corresponding correlations using linear programming \cite{ZKBL1999arxiv} [see Eq.~(4) in the main text].


\begin{small}
\begin{table}[t!]
\begin{center}
\begin{tabular}{ ccccccc } 
\hline
\hline 

Facet class & Name & $L$ & $Q$ (NPA) & $NS$ & $Q/NS$ & $q_{\mathrm{NL}}^{\mathrm{min}}$\\
\hline
1 & \text{CHSH} & 0.0 & 0.2071 & 0.5 & 0.4142 & 0.4142 \\
2 & $I_{[[2,3],[2,2,2]]}$ & 0 & 0.2532 & 0.6667 & 0.3798 & 0.3798\\
3 & $I_{[[2,3],[2,2,2]]}$ & 0.0 & 0.2532 & 0.6667 & 0.3798 & 0.4142\\
4 & $I_{2332}$ & 0.0 & 0.4142 & 1 & 0.4142 & 0.4142\\
5 & $I_{3322}$ & 0.0 & 0.2509 & 1 & 0.2509 & 0.2501\\
6 & $I_{3322}$ & 0.0 & 0.2509 & 1 & 0.2509 & 0.25\\ 
7 & $I_{3322}$ & 0.0 & 0.2509 & 1 & 0.2509 & 0.4142\\
8 & $I_{[[2,2,3],[2,2,2]]}^1$ & 0.0 & 0.4142 & 1 & 0.4142 & 0.4142\\
9&$I_{[[2,2,3],[2,2,2]]}^1$ & 0.0 & 0.4142 & 1 & 0.4142 & 0.4142\\
10&$I_{[[2,2,3],[2,2,2]]}^2$ & 1.0 & 1.3913 & 2 & 0.6956 & 0.3981\\
11 & $I_{[[2,2,3],[2,2,2]]}^2$ &0.0&0.3913& 1 & 0.3913 & 0.3981\\
12 & $I_{[[2,2,3],[2,2,2]]}^2$ &1.0&1.3913& 2 & 0.6956 & 0.3981\\
13 & $I_{[[2,2,3],[2,2,2]]}^2$ &0.0&0.3913& 1 & 0.3913 & 0.3981\\
14 & $I_{[[2,2,3],[2,2,2]]}^1$ &0.0&0.4142& 1 & 0.4142 & 0.4142\\
15 & $I_{[[2,2,3],[2,2,2]]}^3$ &0.0&0.4365& 1.5 & 0.291 & 0.3559\\
16 & $I_{[[2,3,3],[2,2,2]]}^1$ &0.0&0.3015& 0.75 & 0.402 & 0.5981\\
17 & $I_{[[2,3,3],[2,2,2]]}^1$ &0.0&0.3015& 0.75 & 0.402 & 0.5981\\
18 & $I_{[[2,3,3],[2,2,2]]}^2$ &1.0&1.4145& 2 & 0.7073 & 0.4142\\
19 & $I_{[[2,3,3],[2,2,2]]}^2$ &0.0&0.4145& 1 & 0.4145 & 0.4142\\
20 & $I_{[[2,3,3],[2,2,2]]}^3$ &0.0&0.4365& 1.5 & 0.291 & 0.3588\\
21 & $I_{[[2,3,3],[2,2,2]]}^3$ &0.0&0.4365& 1.5 & 0.291 & 0.335\\
22 & $I_{[[2,3,3],[2,2,2]]}^4$ &0.0&0.3913& 1 & 0.3913 & 0.3981\\
23 & $I_{[[2,3,3],[2,2,2]]}^4$ &0.0&0.3913& 1 & 0.3913 & 0.3981\\
24 & $I_{3332}$ & 0.0 & 0.4145 & 1 & 0.4145 & 0.4142\\
\hline 
\hline 
\end{tabular}
\end{center}
\caption{Table providing the local $L$, quantum $Q$, nonsignaling $NS$ bounds for each class of facets of (3,3;3,2). Two measures of the distance of the optimal quantum correlation to the nonsignaling set are also provided, the ratio $Q/NS$ and the nonlocal content $q_{\mathrm{NL}}^{\mathrm{min}}$.}
\label{tab:appA}
\end{table}
\end{small}


\section{Search for FNS=FN=AVN=PT in $(3,3;3,3)$}


We studied the boundary of the quantum correlations for the $(3,3;3,3)$ Bell scenario using the same tools used for $(3,3;3,2)$, namely, computing an upper bound to the maximal quantum violation of facet Bell inequalities and comparing this bound with the maximum value attained by NS correlations. However, the complete list of facets for $(3,3;3,3)$ is not known. A partial list containing 21170 facets was published in \cite{Colbeck:2019PRA}. Here we enlarged this previous list by running PANDA \cite{Lorwald2015} using high performance computing (HPC), using as additional input the known facet inequalities. We obtained a partial list of 8269146 facets that can be found in \cite{Lis}. By computing the tally for each facet, we find that there are at least 4801183 classes in the list. Using HPC, we were able to calculate the quantum bound and NS value for every facet, verifying that in none of the cases there is FNS=FN=AVN=PT.


\bibliographystyle{apsrev4-2}

\begin{thebibliography}{68}%
	\makeatletter
	\providecommand \@ifxundefined [1]{%
		\@ifx{#1\undefined}
	}%
	\providecommand \@ifnum [1]{%
		\ifnum #1\expandafter \@firstoftwo
		\else \expandafter \@secondoftwo
		\fi
	}%
	\providecommand \@ifx [1]{%
		\ifx #1\expandafter \@firstoftwo
		\else \expandafter \@secondoftwo
		\fi
	}%
	\providecommand \natexlab [1]{#1}%
	\providecommand \enquote  [1]{``#1''}%
	\providecommand \bibnamefont  [1]{#1}%
	\providecommand \bibfnamefont [1]{#1}%
	\providecommand \citenamefont [1]{#1}%
	\providecommand \href@noop [0]{\@secondoftwo}%
	\providecommand \href [0]{\begingroup \@sanitize@url \@href}%
	\providecommand \@href[1]{\@@startlink{#1}\@@href}%
	\providecommand \@@href[1]{\endgroup#1\@@endlink}%
	\providecommand \@sanitize@url [0]{\catcode `\\12\catcode `\$12\catcode `\&12\catcode `\#12\catcode `\^12\catcode `\_12\catcode `\%12\relax}%
	\providecommand \@@startlink[1]{}%
	\providecommand \@@endlink[0]{}%
	\providecommand \url  [0]{\begingroup\@sanitize@url \@url }%
	\providecommand \@url [1]{\endgroup\@href {#1}{\urlprefix }}%
	\providecommand \urlprefix  [0]{URL }%
	\providecommand \Eprint [0]{\href }%
	\providecommand \doibase [0]{https://doi.org/}%
	\providecommand \selectlanguage [0]{\@gobble}%
	\providecommand \bibinfo  [0]{\@secondoftwo}%
	\providecommand \bibfield  [0]{\@secondoftwo}%
	\providecommand \translation [1]{[#1]}%
	\providecommand \BibitemOpen [0]{}%
	\providecommand \bibitemStop [0]{}%
	\providecommand \bibitemNoStop [0]{.\EOS\space}%
	\providecommand \EOS [0]{\spacefactor3000\relax}%
	\providecommand \BibitemShut  [1]{\csname bibitem#1\endcsname}%
	\let\auto@bib@innerbib\@empty
	\bibitem [{\citenamefont {Bell}(1964)}]{Bell:1964PHY}%
	\BibitemOpen
	\bibfield  {author} {\bibinfo {author} {\bibfnamefont {J.~S.}\ \bibnamefont {Bell}},\ }\href {https://doi.org/10.1103/PhysicsPhysiqueFizika.1.195} {\bibfield  {journal} {\bibinfo  {journal} {Physics}\ }\textbf {\bibinfo {volume} {1}},\ \bibinfo {pages} {195} (\bibinfo {year} {1964})}\BibitemShut {NoStop}%
	\bibitem [{\citenamefont {Ekert}(1991)}]{Ekert:1991PRL}%
	\BibitemOpen
	\bibfield  {author} {\bibinfo {author} {\bibfnamefont {A.~K.}\ \bibnamefont {Ekert}},\ }\href {https://doi.org/10.1103/PhysRevLett.67.661} {\bibfield  {journal} {\bibinfo  {journal} {Phys. Rev. Lett.}\ }\textbf {\bibinfo {volume} {67}},\ \bibinfo {pages} {661} (\bibinfo {year} {1991})}\BibitemShut {NoStop}%
	\bibitem [{\citenamefont {Colbeck}(2006)}]{Colbeck:2006XXX}%
	\BibitemOpen
	\bibfield  {author} {\bibinfo {author} {\bibfnamefont {R.}~\bibnamefont {Colbeck}},\ }\emph {\bibinfo {title} {Quantum and Relativistic Protocols for Secure Multi-Party Computation}},\ \href@noop {} {Ph.D. thesis},\ \bibinfo  {school} {University of Cambridge} (\bibinfo {year} {2006}),\ \Eprint {https://arxiv.org/abs/0911.3814} {arXiv:0911.3814 [quant-ph]} \BibitemShut {NoStop}%
	\bibitem [{\citenamefont {Brukner}\ \emph {et~al.}(2004)\citenamefont {Brukner}, \citenamefont {{\.Z}ukowski}, \citenamefont {Pan},\ and\ \citenamefont {Zeilinger}}]{Brukner:2004PRL}%
	\BibitemOpen
	\bibfield  {author} {\bibinfo {author} {\bibfnamefont {{\v{C}}.}~\bibnamefont {Brukner}}, \bibinfo {author} {\bibfnamefont {M.}~\bibnamefont {{\.Z}ukowski}}, \bibinfo {author} {\bibfnamefont {J.-W.}\ \bibnamefont {Pan}},\ and\ \bibinfo {author} {\bibfnamefont {A.}~\bibnamefont {Zeilinger}},\ }\href {https://doi.org/10.1103/PhysRevLett.92.127901} {\bibfield  {journal} {\bibinfo  {journal} {Phys. Rev. Lett.}\ }\textbf {\bibinfo {volume} {92}},\ \bibinfo {pages} {127901} (\bibinfo {year} {2004})}\BibitemShut {NoStop}%
	\bibitem [{\citenamefont {Mayers}\ and\ \citenamefont {Yao}(2004)}]{Yao_self}%
	\BibitemOpen
	\bibfield  {author} {\bibinfo {author} {\bibfnamefont {D.}~\bibnamefont {Mayers}}\ and\ \bibinfo {author} {\bibfnamefont {A.}~\bibnamefont {Yao}},\ }\href {http://dl.acm.org/citation.cfm?id=2011827.2011830} {\bibfield  {journal} {\bibinfo  {journal} {Quantum Info. Comput.}\ }\textbf {\bibinfo {volume} {4}},\ \bibinfo {pages} {273} (\bibinfo {year} {2004})}\BibitemShut {NoStop}%
	\bibitem [{\citenamefont {Pitowsky}(1989)}]{Pitowsky:1989}%
	\BibitemOpen
	\bibfield  {author} {\bibinfo {author} {\bibfnamefont {I.}~\bibnamefont {Pitowsky}},\ }\href@noop {} {\emph {\bibinfo {title} {Quantum Probability-Quantum Logic}}},\ Lecture Notes in Physics, Vol.\ 321\ (\bibinfo  {publisher} {Springer-Verlag},\ \bibinfo {address} {Berlin},\ \bibinfo {year} {1989})\BibitemShut {NoStop}%
	\bibitem [{\citenamefont {Cabello}\ \emph {et~al.}(2010)\citenamefont {Cabello}, \citenamefont {Severini},\ and\ \citenamefont {Winter}}]{Cabello:2010XXX}%
	\BibitemOpen
	\bibfield  {author} {\bibinfo {author} {\bibfnamefont {A.}~\bibnamefont {Cabello}}, \bibinfo {author} {\bibfnamefont {S.}~\bibnamefont {Severini}},\ and\ \bibinfo {author} {\bibfnamefont {A.}~\bibnamefont {Winter}},\ }\Eprint {https://arxiv.org/abs/1010.2163} {arXiv:1010.2163 [quant-ph]}  (\bibinfo {year} {2010})\BibitemShut {NoStop}%
	\bibitem [{\citenamefont {Cabello}\ \emph {et~al.}(2014)\citenamefont {Cabello}, \citenamefont {Severini},\ and\ \citenamefont {Winter}}]{CSWPRL2014}%
	\BibitemOpen
	\bibfield  {author} {\bibinfo {author} {\bibfnamefont {A.}~\bibnamefont {Cabello}}, \bibinfo {author} {\bibfnamefont {S.}~\bibnamefont {Severini}},\ and\ \bibinfo {author} {\bibfnamefont {A.}~\bibnamefont {Winter}},\ }\href {https://doi.org/10.1103/PhysRevLett.112.040401} {\bibfield  {journal} {\bibinfo  {journal} {Phys. Rev. Lett.}\ }\textbf {\bibinfo {volume} {112}},\ \bibinfo {pages} {040401} (\bibinfo {year} {2014})}\BibitemShut {NoStop}%
	\bibitem [{\citenamefont {Ramanathan}\ \emph {et~al.}(2016)\citenamefont {Ramanathan}, \citenamefont {Tuziemski}, \citenamefont {Horodecki},\ and\ \citenamefont {Horodecki}}]{Ramanathan:2010PRL}%
	\BibitemOpen
	\bibfield  {author} {\bibinfo {author} {\bibfnamefont {R.}~\bibnamefont {Ramanathan}}, \bibinfo {author} {\bibfnamefont {J.}~\bibnamefont {Tuziemski}}, \bibinfo {author} {\bibfnamefont {M.}~\bibnamefont {Horodecki}},\ and\ \bibinfo {author} {\bibfnamefont {P.}~\bibnamefont {Horodecki}},\ }\href {https://doi.org/10.1103/PhysRevLett.117.050401} {\bibfield  {journal} {\bibinfo  {journal} {Phys. Rev. Lett.}\ }\textbf {\bibinfo {volume} {117}},\ \bibinfo {pages} {050401} (\bibinfo {year} {2016})}\BibitemShut {NoStop}%
	\bibitem [{\citenamefont {Goh}\ \emph {et~al.}(2018)\citenamefont {Goh}, \citenamefont {Kaniewski}, \citenamefont {Wolfe}, \citenamefont {V\'ertesi}, \citenamefont {Wu}, \citenamefont {Cai}, \citenamefont {Liang},\ and\ \citenamefont {Scarani}}]{Goh2018PRA}%
	\BibitemOpen
	\bibfield  {author} {\bibinfo {author} {\bibfnamefont {K.~T.}\ \bibnamefont {Goh}}, \bibinfo {author} {\bibfnamefont {J.~m.~k.}\ \bibnamefont {Kaniewski}}, \bibinfo {author} {\bibfnamefont {E.}~\bibnamefont {Wolfe}}, \bibinfo {author} {\bibfnamefont {T.}~\bibnamefont {V\'ertesi}}, \bibinfo {author} {\bibfnamefont {X.}~\bibnamefont {Wu}}, \bibinfo {author} {\bibfnamefont {Y.}~\bibnamefont {Cai}}, \bibinfo {author} {\bibfnamefont {Y.-C.}\ \bibnamefont {Liang}},\ and\ \bibinfo {author} {\bibfnamefont {V.}~\bibnamefont {Scarani}},\ }\href {https://doi.org/10.1103/PhysRevA.97.022104} {\bibfield  {journal} {\bibinfo  {journal} {Phys. Rev. A}\ }\textbf {\bibinfo {volume} {97}},\ \bibinfo {pages} {022104} (\bibinfo {year} {2018})}\BibitemShut {NoStop}%
	\bibitem [{\citenamefont {Shimony}(1993)}]{Shimony:1993}%
	\BibitemOpen
	\bibfield  {author} {\bibinfo {author} {\bibfnamefont {A.}~\bibnamefont {Shimony}},\ }\href@noop {} {\emph {\bibinfo {title} {Search for a Naturalistic World View. Volume II: Natural Science and Methaphysics}}}\ (\bibinfo  {publisher} {Cambridge University Press},\ \bibinfo {address} {Cambridge, UK},\ \bibinfo {year} {1993})\BibitemShut {NoStop}%
	\bibitem [{\citenamefont {Vieira}\ \emph {et~al.}(2024)\citenamefont {Vieira}, \citenamefont {Ramanathan},\ and\ \citenamefont {Cabello}}]{Vieira:2024X}%
	\BibitemOpen
	\bibfield  {author} {\bibinfo {author} {\bibfnamefont {C.}~\bibnamefont {Vieira}}, \bibinfo {author} {\bibfnamefont {R.}~\bibnamefont {Ramanathan}},\ and\ \bibinfo {author} {\bibfnamefont {A.}~\bibnamefont {Cabello}},\ }\href@noop {} {\bibinfo {title} {Test of the physical significance of {B}ell nonlocality}} (\bibinfo {year} {2024}),\ \Eprint {https://arxiv.org/abs/2402.00801} {arXiv:2402.00801 [quant-ph]} \BibitemShut {NoStop}%
	\bibitem [{\citenamefont {Popescu}\ and\ \citenamefont {Rohrlich}(1994)}]{Popescu:1994FPH}%
	\BibitemOpen
	\bibfield  {author} {\bibinfo {author} {\bibfnamefont {S.}~\bibnamefont {Popescu}}\ and\ \bibinfo {author} {\bibfnamefont {D.}~\bibnamefont {Rohrlich}},\ }\href {https://doi.org/10.1007/BF02058098} {\bibfield  {journal} {\bibinfo  {journal} {Found. Phys.}\ }\textbf {\bibinfo {volume} {24}},\ \bibinfo {pages} {379} (\bibinfo {year} {1994})}\BibitemShut {NoStop}%
	\bibitem [{\citenamefont {Cabello}(2015)}]{Cabello:2015PRL}%
	\BibitemOpen
	\bibfield  {author} {\bibinfo {author} {\bibfnamefont {A.}~\bibnamefont {Cabello}},\ }\href {https://doi.org/10.1103/PhysRevLett.114.220402} {\bibfield  {journal} {\bibinfo  {journal} {Phys. Rev. Lett.}\ }\textbf {\bibinfo {volume} {114}},\ \bibinfo {pages} {220402} (\bibinfo {year} {2015})}\BibitemShut {NoStop}%
	\bibitem [{\citenamefont {Cabello}(2019)}]{Cabello:2019PRA}%
	\BibitemOpen
	\bibfield  {author} {\bibinfo {author} {\bibfnamefont {A.}~\bibnamefont {Cabello}},\ }\href {https://doi.org/10.1103/PhysRevA.100.032120} {\bibfield  {journal} {\bibinfo  {journal} {Phys. Rev. A}\ }\textbf {\bibinfo {volume} {100}},\ \bibinfo {pages} {032120} (\bibinfo {year} {2019})}\BibitemShut {NoStop}%
	\bibitem [{\citenamefont {Gallego}\ \emph {et~al.}(2013)\citenamefont {Gallego}, \citenamefont {Masanes}, \citenamefont {De~la Torre}, \citenamefont {Dhara}, \citenamefont {Aolita},\ and\ \citenamefont {Ac\'{\i}n}}]{Gallego:2013NC}%
	\BibitemOpen
	\bibfield  {author} {\bibinfo {author} {\bibfnamefont {R.}~\bibnamefont {Gallego}}, \bibinfo {author} {\bibfnamefont {L.}~\bibnamefont {Masanes}}, \bibinfo {author} {\bibfnamefont {G.}~\bibnamefont {De~la Torre}}, \bibinfo {author} {\bibfnamefont {C.}~\bibnamefont {Dhara}}, \bibinfo {author} {\bibfnamefont {L.}~\bibnamefont {Aolita}},\ and\ \bibinfo {author} {\bibfnamefont {A.}~\bibnamefont {Ac\'{\i}n}},\ }\href {https://www.nature.com/articles/ncomms3654} {\bibfield  {journal} {\bibinfo  {journal} {Nat. Comm.}\ }\textbf {\bibinfo {volume} {4}},\ \bibinfo {pages} {2654} (\bibinfo {year} {2013})}\BibitemShut {NoStop}%
	\bibitem [{\citenamefont {Clauser}\ \emph {et~al.}(1969)\citenamefont {Clauser}, \citenamefont {Horne}, \citenamefont {Shimony},\ and\ \citenamefont {Holt}}]{Clauser:1969PRL}%
	\BibitemOpen
	\bibfield  {author} {\bibinfo {author} {\bibfnamefont {J.~F.}\ \bibnamefont {Clauser}}, \bibinfo {author} {\bibfnamefont {M.~A.}\ \bibnamefont {Horne}}, \bibinfo {author} {\bibfnamefont {A.}~\bibnamefont {Shimony}},\ and\ \bibinfo {author} {\bibfnamefont {R.~A.}\ \bibnamefont {Holt}},\ }\href {https://doi.org/10.1103/PhysRevLett.23.880} {\bibfield  {journal} {\bibinfo  {journal} {Phys. Rev. Lett.}\ }\textbf {\bibinfo {volume} {23}},\ \bibinfo {pages} {880} (\bibinfo {year} {1969})}\BibitemShut {NoStop}%
	\bibitem [{\citenamefont {Hardy}(1993)}]{HardyPRL1993}%
	\BibitemOpen
	\bibfield  {author} {\bibinfo {author} {\bibfnamefont {L.}~\bibnamefont {Hardy}},\ }\href {https://doi.org/10.1103/PhysRevLett.71.1665} {\bibfield  {journal} {\bibinfo  {journal} {Phys. Rev. Lett.}\ }\textbf {\bibinfo {volume} {71}},\ \bibinfo {pages} {1665} (\bibinfo {year} {1993})}\BibitemShut {NoStop}%
	\bibitem [{\citenamefont {Cabello}(2001{\natexlab{a}})}]{Cabello:2001PRLa}%
	\BibitemOpen
	\bibfield  {author} {\bibinfo {author} {\bibfnamefont {A.}~\bibnamefont {Cabello}},\ }\href {https://doi.org/10.1103/PhysRevLett.86.1911} {\bibfield  {journal} {\bibinfo  {journal} {Phys. Rev. Lett.}\ }\textbf {\bibinfo {volume} {86}},\ \bibinfo {pages} {1911} (\bibinfo {year} {2001}{\natexlab{a}})}\BibitemShut {NoStop}%
	\bibitem [{\citenamefont {Cabello}(2001{\natexlab{b}})}]{Cabello:2001PRLb}%
	\BibitemOpen
	\bibfield  {author} {\bibinfo {author} {\bibfnamefont {A.}~\bibnamefont {Cabello}},\ }\href {https://doi.org/10.1103/PhysRevLett.87.010403} {\bibfield  {journal} {\bibinfo  {journal} {Phys. Rev. Lett.}\ }\textbf {\bibinfo {volume} {87}},\ \bibinfo {pages} {010403} (\bibinfo {year} {2001}{\natexlab{b}})}\BibitemShut {NoStop}%
	\bibitem [{\citenamefont {Aolita}\ \emph {et~al.}(2012)\citenamefont {Aolita}, \citenamefont {Gallego}, \citenamefont {Ac{\'\i}n}, \citenamefont {Chiuri}, \citenamefont {Vallone}, \citenamefont {Mataloni},\ and\ \citenamefont {Cabello}}]{Aolita:2012PRA}%
	\BibitemOpen
	\bibfield  {author} {\bibinfo {author} {\bibfnamefont {L.}~\bibnamefont {Aolita}}, \bibinfo {author} {\bibfnamefont {R.}~\bibnamefont {Gallego}}, \bibinfo {author} {\bibfnamefont {A.}~\bibnamefont {Ac{\'\i}n}}, \bibinfo {author} {\bibfnamefont {A.}~\bibnamefont {Chiuri}}, \bibinfo {author} {\bibfnamefont {G.}~\bibnamefont {Vallone}}, \bibinfo {author} {\bibfnamefont {P.}~\bibnamefont {Mataloni}},\ and\ \bibinfo {author} {\bibfnamefont {A.}~\bibnamefont {Cabello}},\ }\href {https://doi.org/10.1103/PhysRevA.85.032107} {\bibfield  {journal} {\bibinfo  {journal} {Phys. Rev. A}\ }\textbf {\bibinfo {volume} {85}},\ \bibinfo {pages} {032107} (\bibinfo {year} {2012})}\BibitemShut {NoStop}%
	\bibitem [{\citenamefont {Elitzur}\ \emph {et~al.}(1992)\citenamefont {Elitzur}, \citenamefont {Popescu},\ and\ \citenamefont {Rohrlich}}]{Elitzur:1992PLA}%
	\BibitemOpen
	\bibfield  {author} {\bibinfo {author} {\bibfnamefont {A.~C.}\ \bibnamefont {Elitzur}}, \bibinfo {author} {\bibfnamefont {S.}~\bibnamefont {Popescu}},\ and\ \bibinfo {author} {\bibfnamefont {D.}~\bibnamefont {Rohrlich}},\ }\href {https://doi.org/10.1016/0375-9601(92)90952-I} {\bibfield  {journal} {\bibinfo  {journal} {Phys. Lett. A}\ }\textbf {\bibinfo {volume} {162}},\ \bibinfo {pages} {25} (\bibinfo {year} {1992})}\BibitemShut {NoStop}%
	\bibitem [{\citenamefont {Abramsky}\ \emph {et~al.}(2018)\citenamefont {Abramsky}, \citenamefont {Barbosa}, \citenamefont {Car{\`u}}, \citenamefont {{De Silva}}, \citenamefont {Kishida},\ and\ \citenamefont {Mansfield}}]{Abramsky2019Pro}%
	\BibitemOpen
	\bibfield  {author} {\bibinfo {author} {\bibfnamefont {S.}~\bibnamefont {Abramsky}}, \bibinfo {author} {\bibfnamefont {R.}~\bibnamefont {Barbosa}}, \bibinfo {author} {\bibfnamefont {G.}~\bibnamefont {Car{\`u}}}, \bibinfo {author} {\bibfnamefont {N.}~\bibnamefont {{De Silva}}}, \bibinfo {author} {\bibfnamefont {K.}~\bibnamefont {Kishida}},\ and\ \bibinfo {author} {\bibfnamefont {S.}~\bibnamefont {Mansfield}},\ }in\ \href {https://doi.org/10.4230/LIPIcs.TQC.2017.9} {\emph {\bibinfo {booktitle} {12th Conference on the Theory of Quantum Computation, Communication, and Cryptography, TQC 2017}}},\ \bibinfo {series and number} {Leibniz International Proceedings in Informatics, LIPIcs},\ \bibinfo {editor} {edited by\ \bibinfo {editor} {\bibfnamefont {M.}~\bibnamefont {Wilde}}}\ (\bibinfo  {publisher} {Schloss Dagstuhl- Leibniz-Zentrum fur Informatik GmbH, Dagstuhl Publishing},\ \bibinfo {address} {Germany},\ \bibinfo {year} {2018})\ pp.\ \bibinfo {pages} {9:1--9:20}\BibitemShut {NoStop}%
	\bibitem [{\citenamefont {Cubitt}\ \emph {et~al.}(2010)\citenamefont {Cubitt}, \citenamefont {Leung}, \citenamefont {Matthews},\ and\ \citenamefont {Winter}}]{Cubitt:2010PRL}%
	\BibitemOpen
	\bibfield  {author} {\bibinfo {author} {\bibfnamefont {T.~S.}\ \bibnamefont {Cubitt}}, \bibinfo {author} {\bibfnamefont {D.}~\bibnamefont {Leung}}, \bibinfo {author} {\bibfnamefont {W.}~\bibnamefont {Matthews}},\ and\ \bibinfo {author} {\bibfnamefont {A.}~\bibnamefont {Winter}},\ }\href {https://doi.org/10.1103/PhysRevLett.104.230503} {\bibfield  {journal} {\bibinfo  {journal} {Phys. Rev. Lett.}\ }\textbf {\bibinfo {volume} {104}},\ \bibinfo {pages} {230503} (\bibinfo {year} {2010})}\BibitemShut {NoStop}%
	\bibitem [{\citenamefont {Horodecki}\ \emph {et~al.}(2010)\citenamefont {Horodecki}, \citenamefont {Horodecki}, \citenamefont {Horodecki}, \citenamefont {Horodecki}, \citenamefont {Pawlowski},\ and\ \citenamefont {Bourennane}}]{Horodecki:2010XXX}%
	\BibitemOpen
	\bibfield  {author} {\bibinfo {author} {\bibfnamefont {K.}~\bibnamefont {Horodecki}}, \bibinfo {author} {\bibfnamefont {M.}~\bibnamefont {Horodecki}}, \bibinfo {author} {\bibfnamefont {P.}~\bibnamefont {Horodecki}}, \bibinfo {author} {\bibfnamefont {R.}~\bibnamefont {Horodecki}}, \bibinfo {author} {\bibfnamefont {M.}~\bibnamefont {Pawlowski}},\ and\ \bibinfo {author} {\bibfnamefont {M.}~\bibnamefont {Bourennane}},\ }\Eprint {https://arxiv.org/abs/1006.0468} {arXiv:1006.0468 [quant-ph]}  (\bibinfo {year} {2010})\BibitemShut {NoStop}%
	\bibitem [{\citenamefont {Zhen}\ \emph {et~al.}(2023)\citenamefont {Zhen}, \citenamefont {Mao}, \citenamefont {Zhang}, \citenamefont {Xu},\ and\ \citenamefont {Sanders}}]{Zhen:2023PRL}%
	\BibitemOpen
	\bibfield  {author} {\bibinfo {author} {\bibfnamefont {Y.-Z.}\ \bibnamefont {Zhen}}, \bibinfo {author} {\bibfnamefont {Y.}~\bibnamefont {Mao}}, \bibinfo {author} {\bibfnamefont {Y.-Z.}\ \bibnamefont {Zhang}}, \bibinfo {author} {\bibfnamefont {F.}~\bibnamefont {Xu}},\ and\ \bibinfo {author} {\bibfnamefont {B.~C.}\ \bibnamefont {Sanders}},\ }\href {https://doi.org/10.1103/PhysRevLett.131.080801} {\bibfield  {journal} {\bibinfo  {journal} {Phys. Rev. Lett.}\ }\textbf {\bibinfo {volume} {131}},\ \bibinfo {pages} {080801} (\bibinfo {year} {2023})}\BibitemShut {NoStop}%
	\bibitem [{\citenamefont {Vidick}(2017)}]{Vidick:2017XXX}%
	\BibitemOpen
	\bibfield  {author} {\bibinfo {author} {\bibfnamefont {T.}~\bibnamefont {Vidick}},\ }\Eprint {https://arxiv.org/abs/1703.08508} {arXiv:1703.08508 [quant-ph]}  (\bibinfo {year} {2017})\BibitemShut {NoStop}%
	\bibitem [{\citenamefont {Jain}\ \emph {et~al.}(2020)\citenamefont {Jain}, \citenamefont {Miller},\ and\ \citenamefont {Shi}}]{Jain:2020IEE}%
	\BibitemOpen
	\bibfield  {author} {\bibinfo {author} {\bibfnamefont {R.}~\bibnamefont {Jain}}, \bibinfo {author} {\bibfnamefont {C.~A.}\ \bibnamefont {Miller}},\ and\ \bibinfo {author} {\bibfnamefont {Y.}~\bibnamefont {Shi}},\ }\href {https://doi.org/10.1109/TIT.2020.2986740} {\bibfield  {journal} {\bibinfo  {journal} {IEEE Trans. Inf. Theory}\ }\textbf {\bibinfo {volume} {66}},\ \bibinfo {pages} {5567} (\bibinfo {year} {2020})}\BibitemShut {NoStop}%
	\bibitem [{\citenamefont {Einstein}\ \emph {et~al.}(1935)\citenamefont {Einstein}, \citenamefont {Podolsky},\ and\ \citenamefont {Rosen}}]{Einstein:1935PR}%
	\BibitemOpen
	\bibfield  {author} {\bibinfo {author} {\bibfnamefont {A.}~\bibnamefont {Einstein}}, \bibinfo {author} {\bibfnamefont {B.}~\bibnamefont {Podolsky}},\ and\ \bibinfo {author} {\bibfnamefont {N.}~\bibnamefont {Rosen}},\ }\href {https://doi.org/10.1103/PhysRev.47.777} {\bibfield  {journal} {\bibinfo  {journal} {Phys. Rev.}\ }\textbf {\bibinfo {volume} {47}},\ \bibinfo {pages} {777} (\bibinfo {year} {1935})}\BibitemShut {NoStop}%
	\bibitem [{\citenamefont {Stairs}(1983)}]{Stairs:1983PS}%
	\BibitemOpen
	\bibfield  {author} {\bibinfo {author} {\bibfnamefont {A.}~\bibnamefont {Stairs}},\ }\href {https://doi.org/10.1086/289140} {\bibfield  {journal} {\bibinfo  {journal} {Philos. Sci.}\ }\textbf {\bibinfo {volume} {50}},\ \bibinfo {pages} {578} (\bibinfo {year} {1983})}\BibitemShut {NoStop}%
	\bibitem [{\citenamefont {Heywood}\ and\ \citenamefont {Redhead}(1983)}]{HR83}%
	\BibitemOpen
	\bibfield  {author} {\bibinfo {author} {\bibfnamefont {P.}~\bibnamefont {Heywood}}\ and\ \bibinfo {author} {\bibfnamefont {M.~L.~G.}\ \bibnamefont {Redhead}},\ }\href {https://doi.org/10.1007/BF00729511} {\bibfield  {journal} {\bibinfo  {journal} {Found. Phys.}\ }\textbf {\bibinfo {volume} {13}},\ \bibinfo {pages} {481} (\bibinfo {year} {1983})}\BibitemShut {NoStop}%
	\bibitem [{\citenamefont {Kochen}\ and\ \citenamefont {Specker}(1967)}]{Kochen:1967JMM}%
	\BibitemOpen
	\bibfield  {author} {\bibinfo {author} {\bibfnamefont {S.}~\bibnamefont {Kochen}}\ and\ \bibinfo {author} {\bibfnamefont {E.~P.}\ \bibnamefont {Specker}},\ }\href {https://doi.org/10.1512/iumj.1968.17.17004} {\bibfield  {journal} {\bibinfo  {journal} {J. Math. Mech.}\ }\textbf {\bibinfo {volume} {17}},\ \bibinfo {pages} {59} (\bibinfo {year} {1967})}\BibitemShut {NoStop}%
	\bibitem [{\citenamefont {Greenberger}\ \emph {et~al.}(1989)\citenamefont {Greenberger}, \citenamefont {Horne},\ and\ \citenamefont {Zeilinger}}]{GHZ89}%
	\BibitemOpen
	\bibfield  {author} {\bibinfo {author} {\bibfnamefont {D.~M.}\ \bibnamefont {Greenberger}}, \bibinfo {author} {\bibfnamefont {M.~A.}\ \bibnamefont {Horne}},\ and\ \bibinfo {author} {\bibfnamefont {A.}~\bibnamefont {Zeilinger}},\ }in\ \href@noop {} {\emph {\bibinfo {booktitle} {Bell's Theorem, Quantum Theory and Conceptions of the Universe}}}\ (\bibinfo  {publisher} {Springer},\ \bibinfo {year} {1989})\ pp.\ \bibinfo {pages} {69--72}\BibitemShut {NoStop}%
	\bibitem [{\citenamefont {Mermin}(1990{\natexlab{a}})}]{Mermin:1990AJP}%
	\BibitemOpen
	\bibfield  {author} {\bibinfo {author} {\bibfnamefont {N.~D.}\ \bibnamefont {Mermin}},\ }\href {https://doi.org/10.1119/1.16503} {\bibfield  {journal} {\bibinfo  {journal} {Am. J. Phys.}\ }\textbf {\bibinfo {volume} {58}},\ \bibinfo {pages} {731} (\bibinfo {year} {1990}{\natexlab{a}})}\BibitemShut {NoStop}%
	\bibitem [{\citenamefont {Pan}\ \emph {et~al.}(2000)\citenamefont {Pan}, \citenamefont {Bouwmeester}, \citenamefont {Daniell}, \citenamefont {Weinfurter},\ and\ \citenamefont {Zeilinger}}]{Pan:2000NAT}%
	\BibitemOpen
	\bibfield  {author} {\bibinfo {author} {\bibfnamefont {J.-W.}\ \bibnamefont {Pan}}, \bibinfo {author} {\bibfnamefont {D.}~\bibnamefont {Bouwmeester}}, \bibinfo {author} {\bibfnamefont {M.}~\bibnamefont {Daniell}}, \bibinfo {author} {\bibfnamefont {H.}~\bibnamefont {Weinfurter}},\ and\ \bibinfo {author} {\bibfnamefont {A.}~\bibnamefont {Zeilinger}},\ }\href {https://doi.org/10.1038/35000514} {\bibfield  {journal} {\bibinfo  {journal} {Nature (London)}\ }\textbf {\bibinfo {volume} {403}},\ \bibinfo {pages} {515} (\bibinfo {year} {2000})}\BibitemShut {NoStop}%
	\bibitem [{\citenamefont {Leibfried}\ \emph {et~al.}(2004)\citenamefont {Leibfried}, \citenamefont {Barrett}, \citenamefont {Schaetz}, \citenamefont {Britton}, \citenamefont {Chiaverini}, \citenamefont {Itano}, \citenamefont {Jost}, \citenamefont {Langer},\ and\ \citenamefont {Wineland}}]{Leibfried:2004Sci}%
	\BibitemOpen
	\bibfield  {author} {\bibinfo {author} {\bibfnamefont {D.}~\bibnamefont {Leibfried}}, \bibinfo {author} {\bibfnamefont {M.~D.}\ \bibnamefont {Barrett}}, \bibinfo {author} {\bibfnamefont {T.}~\bibnamefont {Schaetz}}, \bibinfo {author} {\bibfnamefont {J.}~\bibnamefont {Britton}}, \bibinfo {author} {\bibfnamefont {J.}~\bibnamefont {Chiaverini}}, \bibinfo {author} {\bibfnamefont {W.~M.}\ \bibnamefont {Itano}}, \bibinfo {author} {\bibfnamefont {J.~D.}\ \bibnamefont {Jost}}, \bibinfo {author} {\bibfnamefont {C.}~\bibnamefont {Langer}},\ and\ \bibinfo {author} {\bibfnamefont {D.~J.}\ \bibnamefont {Wineland}},\ }\href {https://doi.org/10.1126/science.1097576} {\bibfield  {journal} {\bibinfo  {journal} {Science}\ }\textbf {\bibinfo {volume} {304}},\ \bibinfo {pages} {1476} (\bibinfo {year} {2004})}\BibitemShut {NoStop}%
	\bibitem [{\citenamefont {Roos}\ \emph {et~al.}(2004)\citenamefont {Roos}, \citenamefont {Riebe}, \citenamefont {H{\"a}ffner}, \citenamefont {H{\"a}nsel}, \citenamefont {Benhelm}, \citenamefont {Lancaster}, \citenamefont {Becher}, \citenamefont {Schmidt-Kaler},\ and\ \citenamefont {Blatt}}]{Roos:2004Sci}%
	\BibitemOpen
	\bibfield  {author} {\bibinfo {author} {\bibfnamefont {C.~F.}\ \bibnamefont {Roos}}, \bibinfo {author} {\bibfnamefont {M.}~\bibnamefont {Riebe}}, \bibinfo {author} {\bibfnamefont {H.}~\bibnamefont {H{\"a}ffner}}, \bibinfo {author} {\bibfnamefont {W.}~\bibnamefont {H{\"a}nsel}}, \bibinfo {author} {\bibfnamefont {J.}~\bibnamefont {Benhelm}}, \bibinfo {author} {\bibfnamefont {G.~P.~T.}\ \bibnamefont {Lancaster}}, \bibinfo {author} {\bibfnamefont {C.}~\bibnamefont {Becher}}, \bibinfo {author} {\bibfnamefont {F.}~\bibnamefont {Schmidt-Kaler}},\ and\ \bibinfo {author} {\bibfnamefont {R.}~\bibnamefont {Blatt}},\ }\href {https://doi.org/10.1126/science.1097522} {\bibfield  {journal} {\bibinfo  {journal} {Science}\ }\textbf {\bibinfo {volume} {304}},\ \bibinfo {pages} {1478} (\bibinfo {year} {2004})}\BibitemShut {NoStop}%
	\bibitem [{\citenamefont {Cinelli}\ \emph {et~al.}(2005)\citenamefont {Cinelli}, \citenamefont {Barbieri}, \citenamefont {Perris}, \citenamefont {Mataloni},\ and\ \citenamefont {De~Martini}}]{CinelliPRL2005}%
	\BibitemOpen
	\bibfield  {author} {\bibinfo {author} {\bibfnamefont {C.}~\bibnamefont {Cinelli}}, \bibinfo {author} {\bibfnamefont {M.}~\bibnamefont {Barbieri}}, \bibinfo {author} {\bibfnamefont {R.}~\bibnamefont {Perris}}, \bibinfo {author} {\bibfnamefont {P.}~\bibnamefont {Mataloni}},\ and\ \bibinfo {author} {\bibfnamefont {F.}~\bibnamefont {De~Martini}},\ }\href {https://doi.org/10.1103/PhysRevLett.95.240405} {\bibfield  {journal} {\bibinfo  {journal} {Phys. Rev. Lett.}\ }\textbf {\bibinfo {volume} {95}},\ \bibinfo {pages} {240405} (\bibinfo {year} {2005})}\BibitemShut {NoStop}%
	\bibitem [{\citenamefont {Yang}\ \emph {et~al.}(2005)\citenamefont {Yang}, \citenamefont {Zhang}, \citenamefont {Zhang}, \citenamefont {Yin}, \citenamefont {Zhao}, \citenamefont {{\.Z}ukowski}, \citenamefont {Chen},\ and\ \citenamefont {Pan}}]{YangPRL2005}%
	\BibitemOpen
	\bibfield  {author} {\bibinfo {author} {\bibfnamefont {T.}~\bibnamefont {Yang}}, \bibinfo {author} {\bibfnamefont {Q.}~\bibnamefont {Zhang}}, \bibinfo {author} {\bibfnamefont {J.}~\bibnamefont {Zhang}}, \bibinfo {author} {\bibfnamefont {J.}~\bibnamefont {Yin}}, \bibinfo {author} {\bibfnamefont {Z.}~\bibnamefont {Zhao}}, \bibinfo {author} {\bibfnamefont {M.}~\bibnamefont {{\.Z}ukowski}}, \bibinfo {author} {\bibfnamefont {Z.-B.}\ \bibnamefont {Chen}},\ and\ \bibinfo {author} {\bibfnamefont {J.-W.}\ \bibnamefont {Pan}},\ }\href {https://doi.org/10.1103/PhysRevLett.95.240406} {\bibfield  {journal} {\bibinfo  {journal} {Phys. Rev. Lett.}\ }\textbf {\bibinfo {volume} {95}},\ \bibinfo {pages} {240406} (\bibinfo {year} {2005})}\BibitemShut {NoStop}%
	\bibitem [{\citenamefont {Xu}\ \emph {et~al.}(2022)\citenamefont {Xu}, \citenamefont {Zhen}, \citenamefont {Yang}, \citenamefont {Cheng}, \citenamefont {Ren}, \citenamefont {Chen}, \citenamefont {Wang},\ and\ \citenamefont {Wang}}]{Xu:2022PRL}%
	\BibitemOpen
	\bibfield  {author} {\bibinfo {author} {\bibfnamefont {J.-M.}\ \bibnamefont {Xu}}, \bibinfo {author} {\bibfnamefont {Y.-Z.}\ \bibnamefont {Zhen}}, \bibinfo {author} {\bibfnamefont {Y.-X.}\ \bibnamefont {Yang}}, \bibinfo {author} {\bibfnamefont {Z.-M.}\ \bibnamefont {Cheng}}, \bibinfo {author} {\bibfnamefont {Z.-C.}\ \bibnamefont {Ren}}, \bibinfo {author} {\bibfnamefont {K.}~\bibnamefont {Chen}}, \bibinfo {author} {\bibfnamefont {X.-L.}\ \bibnamefont {Wang}},\ and\ \bibinfo {author} {\bibfnamefont {H.-T.}\ \bibnamefont {Wang}},\ }\href {https://doi.org/10.1103/PhysRevLett.129.050402} {\bibfield  {journal} {\bibinfo  {journal} {Phys. Rev. Lett.}\ }\textbf {\bibinfo {volume} {129}},\ \bibinfo {pages} {050402} (\bibinfo {year} {2022})}\BibitemShut {NoStop}%
	\bibitem [{\citenamefont {G\"uhne}\ \emph {et~al.}(2005)\citenamefont {G\"uhne}, \citenamefont {T\'oth}, \citenamefont {Hyllus},\ and\ \citenamefont {Briegel}}]{Guhne:2005PRL}%
	\BibitemOpen
	\bibfield  {author} {\bibinfo {author} {\bibfnamefont {O.}~\bibnamefont {G\"uhne}}, \bibinfo {author} {\bibfnamefont {G.}~\bibnamefont {T\'oth}}, \bibinfo {author} {\bibfnamefont {P.}~\bibnamefont {Hyllus}},\ and\ \bibinfo {author} {\bibfnamefont {H.~J.}\ \bibnamefont {Briegel}},\ }\href {https://doi.org/10.1103/PhysRevLett.95.120405} {\bibfield  {journal} {\bibinfo  {journal} {Phys. Rev. Lett.}\ }\textbf {\bibinfo {volume} {95}},\ \bibinfo {pages} {120405} (\bibinfo {year} {2005})}\BibitemShut {NoStop}%
	\bibitem [{\citenamefont {Mermin}(1990{\natexlab{b}})}]{Mermin:1990PRLa}%
	\BibitemOpen
	\bibfield  {author} {\bibinfo {author} {\bibfnamefont {N.~D.}\ \bibnamefont {Mermin}},\ }\href {https://doi.org/10.1103/PhysRevLett.65.1838} {\bibfield  {journal} {\bibinfo  {journal} {Phys. Rev. Lett.}\ }\textbf {\bibinfo {volume} {65}},\ \bibinfo {pages} {1838} (\bibinfo {year} {1990}{\natexlab{b}})}\BibitemShut {NoStop}%
	\bibitem [{\citenamefont {Cleve}\ \emph {et~al.}(2004)\citenamefont {Cleve}, \citenamefont {{H{\o}yer}}, \citenamefont {Toner},\ and\ \citenamefont {Watrous}}]{CHTW04}%
	\BibitemOpen
	\bibfield  {author} {\bibinfo {author} {\bibfnamefont {R.}~\bibnamefont {Cleve}}, \bibinfo {author} {\bibfnamefont {P.}~\bibnamefont {{H{\o}yer}}}, \bibinfo {author} {\bibfnamefont {B.}~\bibnamefont {Toner}},\ and\ \bibinfo {author} {\bibfnamefont {J.}~\bibnamefont {Watrous}},\ }in\ \href {https://doi.org/10.1109/CCC.2004.1313847} {\emph {\bibinfo {booktitle} {Proceedings. 19th IEEE Annual Conference on Computational Complexity}}}\ (\bibinfo {year} {2004})\ pp.\ \bibinfo {pages} {236--249}\BibitemShut {NoStop}%
	\bibitem [{\citenamefont {Aravind}(2004)}]{Aravind:2004AJP}%
	\BibitemOpen
	\bibfield  {author} {\bibinfo {author} {\bibfnamefont {P.~K.}\ \bibnamefont {Aravind}},\ }\href {https://doi.org/10.1119/1.1773173} {\bibfield  {journal} {\bibinfo  {journal} {Am. J. Phys.}\ }\textbf {\bibinfo {volume} {72}},\ \bibinfo {pages} {1303} (\bibinfo {year} {2004})}\BibitemShut {NoStop}%
	\bibitem [{\citenamefont {Brassard}\ \emph {et~al.}(2005{\natexlab{a}})\citenamefont {Brassard}, \citenamefont {Broadbent},\ and\ \citenamefont {Tapp}}]{GBT05}%
	\BibitemOpen
	\bibfield  {author} {\bibinfo {author} {\bibfnamefont {G.}~\bibnamefont {Brassard}}, \bibinfo {author} {\bibfnamefont {A.}~\bibnamefont {Broadbent}},\ and\ \bibinfo {author} {\bibfnamefont {A.}~\bibnamefont {Tapp}},\ }\href {https://doi.org/10.1007/s10701-005-7353-4} {\bibfield  {journal} {\bibinfo  {journal} {Found. Phys.}\ }\textbf {\bibinfo {volume} {35}},\ \bibinfo {pages} {1877} (\bibinfo {year} {2005}{\natexlab{a}})}\BibitemShut {NoStop}%
	\bibitem [{\citenamefont {Broadbent}(2008)}]{BroadbentPhD2008}%
	\BibitemOpen
	\bibfield  {author} {\bibinfo {author} {\bibfnamefont {A.~L.}\ \bibnamefont {Broadbent}},\ }\emph {\bibinfo {title} {Quantum Nonlocality, Cryptography and Complexity}},\ \href {https://papyrus.bib.umontreal.ca/xmlui/handle/1866/6448} {Ph.D. thesis},\ \bibinfo  {school} {Universit\'e de Montr\'eal} (\bibinfo {year} {2008})\BibitemShut {NoStop}%
	\bibitem [{\citenamefont {Bravyi}\ \emph {et~al.}(2018)\citenamefont {Bravyi}, \citenamefont {Gosset},\ and\ \citenamefont {K{\"o}nig}}]{Bravyi:2018SCI}%
	\BibitemOpen
	\bibfield  {author} {\bibinfo {author} {\bibfnamefont {S.}~\bibnamefont {Bravyi}}, \bibinfo {author} {\bibfnamefont {D.}~\bibnamefont {Gosset}},\ and\ \bibinfo {author} {\bibfnamefont {R.}~\bibnamefont {K{\"o}nig}},\ }\href {https://doi.org/10.1126/science.aar3106} {\bibfield  {journal} {\bibinfo  {journal} {Science}\ }\textbf {\bibinfo {volume} {362}},\ \bibinfo {pages} {308} (\bibinfo {year} {2018})}\BibitemShut {NoStop}%
	\bibitem [{\citenamefont {Ji}\ \emph {et~al.}(2021)\citenamefont {Ji}, \citenamefont {Natarajan}, \citenamefont {Vidick}, \citenamefont {Wright},\ and\ \citenamefont {Yuen}}]{Ji:2021CACM}%
	\BibitemOpen
	\bibfield  {author} {\bibinfo {author} {\bibfnamefont {Z.}~\bibnamefont {Ji}}, \bibinfo {author} {\bibfnamefont {A.}~\bibnamefont {Natarajan}}, \bibinfo {author} {\bibfnamefont {T.}~\bibnamefont {Vidick}}, \bibinfo {author} {\bibfnamefont {J.}~\bibnamefont {Wright}},\ and\ \bibinfo {author} {\bibfnamefont {H.}~\bibnamefont {Yuen}},\ }\href {https://doi.org/10.1145/3485628} {\bibfield  {journal} {\bibinfo  {journal} {Comm. ACM}\ }\textbf {\bibinfo {volume} {64}},\ \bibinfo {pages} {131} (\bibinfo {year} {2021})}\BibitemShut {NoStop}%
	\bibitem [{\citenamefont {Fu}\ and\ \citenamefont {Miller}(2018)}]{Fu:2018PRA}%
	\BibitemOpen
	\bibfield  {author} {\bibinfo {author} {\bibfnamefont {H.}~\bibnamefont {Fu}}\ and\ \bibinfo {author} {\bibfnamefont {C.~A.}\ \bibnamefont {Miller}},\ }\href {https://doi.org/10.1103/PhysRevA.97.032324} {\bibfield  {journal} {\bibinfo  {journal} {Phys. Rev. A}\ }\textbf {\bibinfo {volume} {97}},\ \bibinfo {pages} {032324} (\bibinfo {year} {2018})}\BibitemShut {NoStop}%
	\bibitem [{\citenamefont {Leditzky}\ \emph {et~al.}(2020)\citenamefont {Leditzky}, \citenamefont {Alhejji}, \citenamefont {Levin},\ and\ \citenamefont {Smith}}]{LALS2020NATCOM}%
	\BibitemOpen
	\bibfield  {author} {\bibinfo {author} {\bibfnamefont {F.}~\bibnamefont {Leditzky}}, \bibinfo {author} {\bibfnamefont {M.~A.}\ \bibnamefont {Alhejji}}, \bibinfo {author} {\bibfnamefont {J.}~\bibnamefont {Levin}},\ and\ \bibinfo {author} {\bibfnamefont {G.}~\bibnamefont {Smith}},\ }\href {https://www.nature.com/articles/s41467-020-15240-w} {\bibfield  {journal} {\bibinfo  {journal} {Nat. Commun.}\ }\textbf {\bibinfo {volume} {11}},\ \bibinfo {pages} {1497} (\bibinfo {year} {2020})}\BibitemShut {NoStop}%
	\bibitem [{\citenamefont {Cope}\ and\ \citenamefont {Colbeck}(2019)}]{Colbeck:2019PRA}%
	\BibitemOpen
	\bibfield  {author} {\bibinfo {author} {\bibfnamefont {T.}~\bibnamefont {Cope}}\ and\ \bibinfo {author} {\bibfnamefont {R.}~\bibnamefont {Colbeck}},\ }\href {https://doi.org/10.1103/PhysRevA.100.022114} {\bibfield  {journal} {\bibinfo  {journal} {Phys. Rev. A}\ }\textbf {\bibinfo {volume} {100}},\ \bibinfo {pages} {022114} (\bibinfo {year} {2019})}\BibitemShut {NoStop}%
	\bibitem [{\citenamefont {Matouek}\ and\ \citenamefont {G\"{a}rtner}(2006)}]{Matouek:2006}%
	\BibitemOpen
	\bibfield  {author} {\bibinfo {author} {\bibfnamefont {J.}~\bibnamefont {Matouek}}\ and\ \bibinfo {author} {\bibfnamefont {B.}~\bibnamefont {G\"{a}rtner}},\ }\href@noop {} {\emph {\bibinfo {title} {Understanding and Using Linear Programming}}}\ (\bibinfo  {publisher} {Springer-Verlag},\ \bibinfo {address} {Berlin, Heidelberg},\ \bibinfo {year} {2006})\BibitemShut {NoStop}%
	\bibitem [{\citenamefont {Gisin}\ \emph {et~al.}(2007)\citenamefont {Gisin}, \citenamefont {M\'ethot},\ and\ \citenamefont {Scarani}}]{Gisin:2007IJQI}%
	\BibitemOpen
	\bibfield  {author} {\bibinfo {author} {\bibfnamefont {N.}~\bibnamefont {Gisin}}, \bibinfo {author} {\bibfnamefont {A.~A.}\ \bibnamefont {M\'ethot}},\ and\ \bibinfo {author} {\bibfnamefont {V.}~\bibnamefont {Scarani}},\ }\href {https://doi.org/10.1142/S021974990700289X} {\bibfield  {journal} {\bibinfo  {journal} {Int. J. Quant. Inf.}\ }\textbf {\bibinfo {volume} {5}},\ \bibinfo {pages} {525} (\bibinfo {year} {2007})}\BibitemShut {NoStop}%
	\bibitem [{\citenamefont {Brassard}\ \emph {et~al.}(2005{\natexlab{b}})\citenamefont {Brassard}, \citenamefont {M\'ethot},\ and\ \citenamefont {Tapp}}]{Brassard:2005}%
	\BibitemOpen
	\bibfield  {author} {\bibinfo {author} {\bibfnamefont {G.}~\bibnamefont {Brassard}}, \bibinfo {author} {\bibfnamefont {A.~A.}\ \bibnamefont {M\'ethot}},\ and\ \bibinfo {author} {\bibfnamefont {A.}~\bibnamefont {Tapp}},\ }\href {https://doi.org/10.26421/QIC5.45-2} {\bibfield  {journal} {\bibinfo  {journal} {Quantum Inf. Comput.}\ }\textbf {\bibinfo {volume} {5}},\ \bibinfo {pages} {275} (\bibinfo {year} {2005}{\natexlab{b}})}\BibitemShut {NoStop}%
	\bibitem [{\citenamefont {Barrett}\ \emph {et~al.}(2006)\citenamefont {Barrett}, \citenamefont {Kent},\ and\ \citenamefont {Pironio}}]{Barrett:2006PRL}%
	\BibitemOpen
	\bibfield  {author} {\bibinfo {author} {\bibfnamefont {J.}~\bibnamefont {Barrett}}, \bibinfo {author} {\bibfnamefont {A.}~\bibnamefont {Kent}},\ and\ \bibinfo {author} {\bibfnamefont {S.}~\bibnamefont {Pironio}},\ }\href {https://doi.org/10.1103/PhysRevLett.97.170409} {\bibfield  {journal} {\bibinfo  {journal} {Phys. Rev. Lett.}\ }\textbf {\bibinfo {volume} {97}},\ \bibinfo {pages} {170409} (\bibinfo {year} {2006})}\BibitemShut {NoStop}%
	\bibitem [{\citenamefont {Jesus}\ and\ \citenamefont {Zambrini~Cruzeiro}(2023)}]{Zambrini:2022XXX}%
	\BibitemOpen
	\bibfield  {author} {\bibinfo {author} {\bibfnamefont {J.}~\bibnamefont {Jesus}}\ and\ \bibinfo {author} {\bibfnamefont {E.}~\bibnamefont {Zambrini~Cruzeiro}},\ }\href {https://doi.org/10.1103/PhysRevA.108.052220} {\bibfield  {journal} {\bibinfo  {journal} {Phys. Rev. A}\ }\textbf {\bibinfo {volume} {108}},\ \bibinfo {pages} {052220} (\bibinfo {year} {2023})}\BibitemShut {NoStop}%
	\bibitem [{\citenamefont {Navascu{\'e}s}\ \emph {et~al.}(2008)\citenamefont {Navascu{\'e}s}, \citenamefont {Pironio},\ and\ \citenamefont {Ac{\'\i}n}}]{NPA_NJP}%
	\BibitemOpen
	\bibfield  {author} {\bibinfo {author} {\bibfnamefont {M.}~\bibnamefont {Navascu{\'e}s}}, \bibinfo {author} {\bibfnamefont {S.}~\bibnamefont {Pironio}},\ and\ \bibinfo {author} {\bibfnamefont {A.}~\bibnamefont {Ac{\'\i}n}},\ }\href {https://doi.org/10.1088/1367-2630/10/7/073013} {\bibfield  {journal} {\bibinfo  {journal} {New J. Phys.}\ }\textbf {\bibinfo {volume} {10}},\ \bibinfo {pages} {073013} (\bibinfo {year} {2008})}\BibitemShut {NoStop}%
	\bibitem [{\citenamefont {Amselem}\ \emph {et~al.}(2012)\citenamefont {Amselem}, \citenamefont {Danielsen}, \citenamefont {L\'opez-Tarrida}, \citenamefont {Portillo}, \citenamefont {Bourennane},\ and\ \citenamefont {Cabello}}]{Amselem:2012PRL}%
	\BibitemOpen
	\bibfield  {author} {\bibinfo {author} {\bibfnamefont {E.}~\bibnamefont {Amselem}}, \bibinfo {author} {\bibfnamefont {L.~E.}\ \bibnamefont {Danielsen}}, \bibinfo {author} {\bibfnamefont {A.~J.}\ \bibnamefont {L\'opez-Tarrida}}, \bibinfo {author} {\bibfnamefont {J.~R.}\ \bibnamefont {Portillo}}, \bibinfo {author} {\bibfnamefont {M.}~\bibnamefont {Bourennane}},\ and\ \bibinfo {author} {\bibfnamefont {A.}~\bibnamefont {Cabello}},\ }\href {https://doi.org/10.1103/PhysRevLett.108.200405} {\bibfield  {journal} {\bibinfo  {journal} {Phys. Rev. Lett.}\ }\textbf {\bibinfo {volume} {108}},\ \bibinfo {pages} {200405} (\bibinfo {year} {2012})}\BibitemShut {NoStop}%
	\bibitem [{\citenamefont {Mančinska}(2014)}]{Mancinska}%
	\BibitemOpen
	\bibfield  {author} {\bibinfo {author} {\bibfnamefont {L.}~\bibnamefont {Mančinska}},\ }\bibinfo {title} {Maximally entangled states in pseudo-telepathy games},\ in\ \href {https://doi.org/10.1007/978-3-319-13350-8} {\emph {\bibinfo {booktitle} {Computing with New Resources}}},\ \bibinfo {series} {Lecture Notes in Computer Science}, Vol.\ \bibinfo {volume} {8808}\ (\bibinfo {year} {2014})\ pp.\ \bibinfo {pages} {200--207}\BibitemShut {NoStop}%
	\bibitem [{\citenamefont {Chen}\ \emph {et~al.}(2003)\citenamefont {Chen}, \citenamefont {Pan}, \citenamefont {Zhang}, \citenamefont {Brukner},\ and\ \citenamefont {Zeilinger}}]{ChenPRL2003}%
	\BibitemOpen
	\bibfield  {author} {\bibinfo {author} {\bibfnamefont {Z.-B.}\ \bibnamefont {Chen}}, \bibinfo {author} {\bibfnamefont {J.-W.}\ \bibnamefont {Pan}}, \bibinfo {author} {\bibfnamefont {Y.-D.}\ \bibnamefont {Zhang}}, \bibinfo {author} {\bibfnamefont {{\v{C}}.}~\bibnamefont {Brukner}},\ and\ \bibinfo {author} {\bibfnamefont {A.}~\bibnamefont {Zeilinger}},\ }\href {https://doi.org/10.1103/PhysRevLett.90.160408} {\bibfield  {journal} {\bibinfo  {journal} {Phys. Rev. Lett.}\ }\textbf {\bibinfo {volume} {90}},\ \bibinfo {pages} {160408} (\bibinfo {year} {2003})}\BibitemShut {NoStop}%
	\bibitem [{\citenamefont {Pironio}(2005)}]{pironio2005lifting}%
	\BibitemOpen
	\bibfield  {author} {\bibinfo {author} {\bibfnamefont {S.}~\bibnamefont {Pironio}},\ }\href {https://doi.org/10.1063/1.1928727} {\bibfield  {journal} {\bibinfo  {journal} {J. Math. Phys.}\ }\textbf {\bibinfo {volume} {46}},\ \bibinfo {pages} {062112} (\bibinfo {year} {2005})}\BibitemShut {NoStop}%
	\bibitem [{\citenamefont {Collins}\ and\ \citenamefont {Gisin}(2004)}]{Collins:2004JPA}%
	\BibitemOpen
	\bibfield  {author} {\bibinfo {author} {\bibfnamefont {D.}~\bibnamefont {Collins}}\ and\ \bibinfo {author} {\bibfnamefont {N.}~\bibnamefont {Gisin}},\ }\href {https://iopscience.iop.org/article/10.1088/0305-4470/37/5/021/meta} {\bibfield  {journal} {\bibinfo  {journal} {J. Phys. A: Math. Gen.}\ }\textbf {\bibinfo {volume} {37}},\ \bibinfo {pages} {1775} (\bibinfo {year} {2004})}\BibitemShut {NoStop}%
	\bibitem [{\citenamefont {East}\ and\ \citenamefont {Mitchell}(2021)}]{East:2021NZJM}%
	\BibitemOpen
	\bibfield  {author} {\bibinfo {author} {\bibfnamefont {J.}~\bibnamefont {East}}\ and\ \bibinfo {author} {\bibfnamefont {J.}~\bibnamefont {Mitchell}},\ }\href {https://doi.org/10.53733/108} {\bibfield  {journal} {\bibinfo  {journal} {N. Z. J. Math.}\ }\textbf {\bibinfo {volume} {51}},\ \bibinfo {pages} {85} (\bibinfo {year} {2021})}\BibitemShut {NoStop}%
	\bibitem [{CVX()}]{CVX}%
	\BibitemOpen
	\href@noop {} {}\bibinfo {howpublished} {\url{http://cvxr.com/cvx/}}\BibitemShut {NoStop}%
	\bibitem [{SDP()}]{SDPT3}%
	\BibitemOpen
	\href@noop {} {}\bibinfo {howpublished} {\url{https://github.com/sqlp/sdpt3}}\BibitemShut {NoStop}%
	\bibitem [{\citenamefont {{\.Z}ukowski}\ \emph {et~al.}(1999)\citenamefont {{\.Z}ukowski}, \citenamefont {Kaszlikowski}, \citenamefont {Baturo},\ and\ \citenamefont {Larsson}}]{ZKBL1999arxiv}%
	\BibitemOpen
	\bibfield  {author} {\bibinfo {author} {\bibfnamefont {M.}~\bibnamefont {{\.Z}ukowski}}, \bibinfo {author} {\bibfnamefont {D.}~\bibnamefont {Kaszlikowski}}, \bibinfo {author} {\bibfnamefont {A.}~\bibnamefont {Baturo}},\ and\ \bibinfo {author} {\bibfnamefont {J.-{\AA}.}\ \bibnamefont {Larsson}},\ }\Eprint {https://arxiv.org/abs/quant-ph/9910058} {arXiv:quant-ph/9910058 [quant-ph]}  (\bibinfo {year} {1999})\BibitemShut {NoStop}%
	\bibitem [{\citenamefont {L{\"o}rwald}\ and\ \citenamefont {Reinelt}(2015)}]{Lorwald2015}%
	\BibitemOpen
	\bibfield  {author} {\bibinfo {author} {\bibfnamefont {S.}~\bibnamefont {L{\"o}rwald}}\ and\ \bibinfo {author} {\bibfnamefont {G.}~\bibnamefont {Reinelt}},\ }\href {https://doi.org/10.1007/s13675-015-0040-0} {\bibfield  {journal} {\bibinfo  {journal} {EURO J. Comput. Optim.}\ }\textbf {\bibinfo {volume} {3}},\ \bibinfo {pages} {297} (\bibinfo {year} {2015})}\BibitemShut {NoStop}%
	\bibitem [{Lis()}]{Lis}%
	\BibitemOpen
	\href@noop {} {}\bibinfo {howpublished} {\url{https://1drv.ms/f/s!AsQvM75Z11kn2dkRU_5KnJcz3WHtMg?e=dhgiV8}}\BibitemShut {NoStop}%
\end{thebibliography}

%


\end{document}